\documentclass[twocolumn,showpacs,preprintnumbers,amssymb,nofootinbib, aps,prd, eqsecnum]{revtex4-1}
\usepackage{graphicx, booktabs, epsf, epsfig}
\usepackage{bm} 
\usepackage{amsmath}
\usepackage{mathpazo}

\usepackage{color}
\usepackage{ulem} 
\usepackage{float}
\usepackage{url}

\definecolor{green}{rgb}{0.1,0.6,0.1}

\def\be{\begin{equation}}
\def\ee{\end{equation}}
\def\beq{\begin{eqnarray}}
\def\eeq{\end{eqnarray}}
\def \bsp{\begin{split}}
\def \ensp{ \end{split} }

\def\f{\frac}

\begin{document}

\title{Prospects for estimating parameters from gravitational waves of superspinar binaries}

\author{Nami Uchikata$^1$}\email{uchikata@icrr.u-tokyo.ac.jp}

\author{Tatsuya Narikawa$^1$}\email{narikawa@icrr.u-tokyo.ac.jp}

\affiliation{
$^{1}$Institute for Cosmic Ray Research, University of Tokyo, Kashiwa City, Chiba 277-8582, Japan\\
}
\date{\today} 

\begin{abstract}
To date, close to fifty black hole binary mergers were observed by the LIGO and Virgo detectors.  
The analyses have been done with an assumption that these objects are black holes by limiting the spin prior to the Kerr bound.
However, the above assumption is not valid for superspinars, which have the Kerr geometry but rotate beyond the Kerr bound. 
In this study, we investigate whether and how the limited spin prior range causes a bias in parameter estimation for superspinars if they are detected.
To this end, we estimate binary parameters of the simulated inspiral signals of the gravitational waves of compact binaries by assuming that at least one component of them is a superspinar.
We have found that when the primary is a superspinar, both mass and spin parameters are biased in parameter estimation due to the limited spin prior range.
In this case, the extended prior range is strongly favored compared to the limited one. 
On the other hand, when the primary is a black hole, we do not see much bias in parameter estimation due to the limited spin prior range, even though the secondary is a superspinar.
We also apply the analysis to black hole binary merger events GW170608 and GW190814, which have a long and loud inspiral signal.
We do not see any preference of superspinars from the model selection for both events.
We conclude that the extension of the spin prior range is necessary for accurate parameter estimation if highly spinning primary objects are found, while it is difficult to identify superspinars if they are only the secondary objects.
Nevertheless, the bias in parameter estimation of spin for the limited spin prior range can be a clue of the existence of superspinars.

\end{abstract}


\maketitle

\newpage
\section{Introduction}
Kerr black hole is a unique solution in an asymptotically-flat stationary and axisymmetric vacuum spacetime \cite{Carter:1971zc}.
It is characterized by two physical parameters: mass $m$ and angular momentum $J$.
The angular momentum is bounded as $c J/ (G m^2)   \le 1$ so that the spacetime singularity and closed time like curves are hidden by the event horizon.  
Here, $c$ and $G$ are the speed of light and the gravitational constant.
However, the Kerr bound $c J/ (G m^2) = 1$ is no longer necessary in string theory and the Kerr geometry with $c J/ (G m^2) > 1$ is named as a superspinar \cite{Gimon:2007ur}. 
Superspinar has been proposed as a source candidate of high energy cosmic rays because of the large efficiency of the energy extraction (See also, \cite{Patil:2015fua}).
It is assumed that the singularity and closed timelike curves can be modified by the stringy effect .

To assume superspinars as  possible astrophysical compact objects, their stability should be confirmed.
From the analysis of linear perturbations in the spacetime, some studies show that superspinars are unstable \cite{Cardoso:2008kj,Pani:2010jz}, while some show that they can be stable \cite{Nakao:2017rgv,Roy:2019uuy}.
This is because we do not know the physically appropriate boundary conditions to solve equations of perturbations in the superspinning Kerr spacetime.
The above studies show that the stability depends on the boundary conditions in the vicinity of the singularity.
Moreover, horizonless highly spinning objects are unstable due to the ergoregion \cite{Friedman:1978hf,Friedman:1978wla,Maggio:2018ivz}.
However, again we do not know the exact physics near the singularity, the ergoregion instability is unknown for superspinars.
Here, we assume that superspinars exist stably at least for the timescale of the binary evolution in the detector sensitive band of current ground-based gravitational wave detectors.

In parameter estimation for binary black holes, we normally assume that the objects are black holes \cite{TheLIGOScientific:2016wfe,170104,170814, 170608,gwtc-1,190412, 190521}, that is, we limit the spin prior range up to the Kerr bound. 
Because of this limitation, we might misidentify superspinars as highly spinning or extremal black holes even if they are detected.
To date, some detected binary black hole mergers show the possibility of large spins close to the Kerr bound \cite{190521, Abbott:2020mjq,Abbott:2020niy}, and a different analysis method also shows the possible existence of extreme black holes \cite{Biscoveanu:2020are}.
In addition, different spin priors lead to different properties of binaries as argued for GW190412 \cite{Mandel:2020lhv,Zevin:2020gxf}.
In this study, we investigate whether and how the spin prior limited by the Kerr bound affects parameter estimation assuming that superspinars are detected as black holes.
Since there is no proper waveform model for binary mergers of superspinars particularly in the post-inspiral phase, we focus on the inspiral part of the binary and use the TalylorF2 waveform model.
The TaylorF2 waveform model is an inspiral waveform model in the frequency domain, obtained by the post-Newtonian (PN) expansion \cite{Dhurandhar:1992mw,Buonanno:2009zt,Blanchet:2013haa, Khan:2015jqa}. 
That is, there is no restriction of the magnitude of the spin in this waveform model.
The detectability of the spin magnitude larger than the Kerr bound using the TaylorF2 waveform model by Fisher analysis is shown in Refs.~\cite{VanDenBroeck:2006ar, Wade:2013hoa}.
The detection of superspinars in extreme mass ratio inspirals by a space-based gravitational wave detector is discussed  in \cite{Piovano:2020ooe,Piovano:2020zin}. 

In this paper, we perform Bayesian analysis for gravitational wave signals from superspinar binaries for the first time.
We analyze the TaylorF2 model waveform by assuming at least one component of the binary is a superspinar and perform Bayesian parameter estimation with two spin priors; one is restricted to the Kerr bound and the other is extended beyond the bound.
We also apply the analysis to black hole binary events GW170608 \cite{170608} and GW190814 \cite{Abbott:2020khf}, which have a long and loud inspiral signal \cite{Abbott:2020jks}.

This paper is organized as follows. 
In Sec.~II, we summarize the basic properties of superspinars including circular orbits in the spacetime.
In Sec.~III, we explain the method and settings for Bayesian parameter estimation.
In Sec.~IV, we show the posteriors of parameter estimation of injection study and also compare the evidences from two priors.
We also discuss the bias of estimated mass parameters observed in the limited spin prior case.
We further show the results for GW170608 and GW190814 using the extended spin prior range.
We summarize and conclude our study in Sec.~V.

\section{Superspinars} 
In this section, we briefly explain the spacetime structure of superspinars.
We use geometrical units $c=G=1$.

\subsection{Metric} 
Kerr metric in the Boyer-Lindquist coordinate is 
\be
\begin{split}
ds^2 = &-dt ^2 + \Sigma \left ( \f{dr^2}{\Delta} + d \theta^2\right )  \\
& + \f{2Mr}{\Sigma}\left( a \sin \theta d \phi -dt \right )^2 + (r^2 + a^2 )\sin^2 \theta d \phi^2,
\label{metric}
\end{split}
\ee
where $\Delta = r^2 -2 m r + a^2$ and $\Sigma = r^2 + a^2 \cos^2 \theta$.
Parameters $m$ and $a$ are black hole mass and its spin.
For $a^2 \le m^2$, the event horizon exist at $r_+ = m + \sqrt{m^2- a^2}$.
Superspinar represents a spacetime geometry described by the Kerr metric Eq.~\eqref{metric} with $\chi >1$, where $\chi$ is a dimensionless spin defined as $\chi = a/m$.
We assume that the spacetime singularity and closed timelike curves are modified by the stringy effect \cite{Gimon:2007ur} and the details of the mechanism are beyond our scope.

\subsection{Orbits} 
\label{sec-orbit}
\begin{figure}[t]
\includegraphics[scale=0.35,trim=20 160 20 160 ]{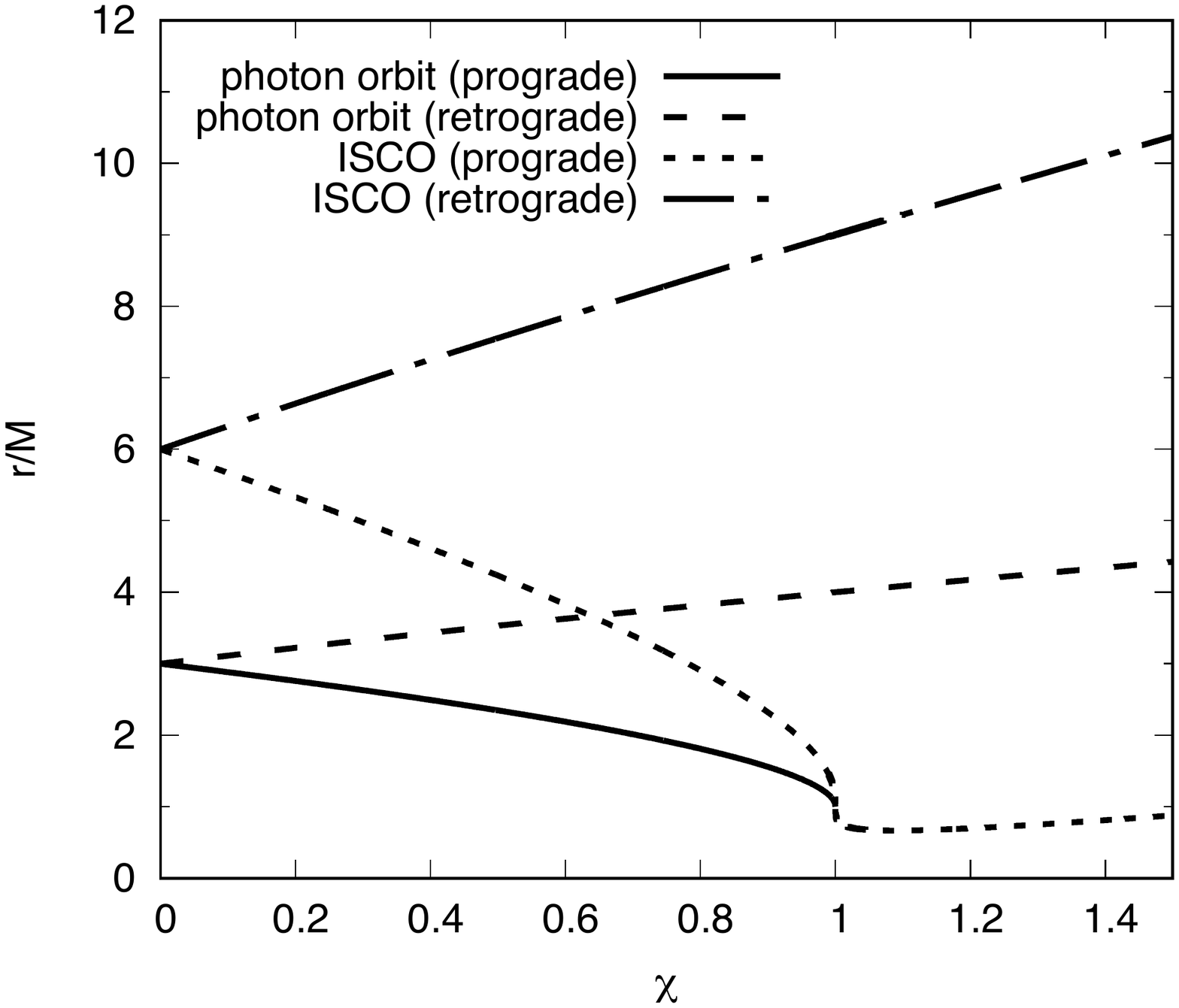}
\caption{Radii of photon orbits and the ISCOs as a function of the dimensionless spin.
There is no prograde photon orbit for $\chi > 1$.}
\label{orbits}
\end{figure}

In this subsection, we summarize the radii of circular orbits in the equatorial plane in the superspinning Kerr spacetime.
We can obtain them from the radial equation of motion in the spacetime.
These orbits and their derivations for the Kerr case ($\chi \le 1$) are summarized in \cite{Bardeen:1972fi}.
For the superspinning case, the orbits are also shown in \cite{Bambi:2008jg,Zhou:2019kwb}.
Photon orbits are unstable orbits for massless objects, whose radius is given as
\be
\f{r_{\rm ph}}{m} =
\begin{cases}
& 2 \left \{ 1 + \cos \left [ \f{2}{3} \cos^{-1}\left ( \mp \chi \right ) \right ] \right \}, \quad (\chi \le 1) \\
& 2 + \f{1}{F(\chi)} +  F(\chi) \quad (\chi >1),
\end{cases}
\ee
where 
\be
F(\chi) =\left ( 2 \chi^2 + 2 \chi \sqrt{\chi^2 -1} -1 \right )^{1/3}.
\ee
The minus (plus) sign corresponds to prograde (retrograde) orbits against the black hole's or superspinar's rotation.
For $\chi >1 $, there is no prograde photon orbit.
The innermost stable circular orbits (ISCOs) are orbits for massive objects, beyond which orbiting objects will fall into the central black hole or superspinar.
Therefore, we usually assume the gravitational wave frequency at the ISCO as the end of the inspiral phase for binary mergers. 
The radius of the orbits can be given as
\be
\begin{split}
\f{r_{\rm ISCO}}{m} 
& = 3 + H(\chi) \\
& \mp \sqrt{ 6+ 2 \chi^2 - \f{ \chi^4 - 10 \chi^2 + 9}{ G(\chi)}  -G(\chi) +\f{16 \chi^2}{H(\chi)}  },
\end{split}
\ee
where 
\be
\begin{split}
G(\chi) & =\left ( 27- 45\chi^2 + 17 \chi^4 + \chi^6 + 8 \chi^3|\chi^2 - 1| \right )^{1/3}, \\
H(\chi) & =  \sqrt{3+ \chi^2 +\f{ \chi^4 - 10 \chi^2 + 9}{ G(\chi)} + G(\chi)  }.
\end{split}
\ee
Again minus/plus sign corresponds to prograde/retrograde orbits against the black hole's or superspinar's rotation.
Rough estimates for the gravitational wave frequency at the prograde ISCO $f_{\rm ISCO}$ are $(r_{\rm ISCO}, f_{\rm ISCO}) \approx [0.9 m, 75760(M_{\odot}/m) {\rm Hz}]$, $[0.8 m, 90293(M_{\odot}/m) {\rm Hz}]$.
For the retrograde case, $(r_{\rm ISCO}, f_{\rm ISCO}) \approx [9.01 m, 2389(M_{\odot}/m) {\rm Hz}]$.
Here, we only consider the leading order term in the phase of the gravitational waveform.

The radii of orbits are shown in Fig.~\ref{orbits} as functions of a dimensionless spin.
The radius of retrograde orbits is a increasing function of $\chi$.
On the other hand, the radius of prograde photon orbits is a decreasing function of $\chi$, and it reaches $M$ at $\chi=1$.
Beyond $\chi =1$, there is no prograde photon orbit.
The radius of prograde ISCO decreases as $\chi $ increases and it becomes smaller than $M$ for $\chi >1$.
However, it slightly increases and becomes larger than $M$ for $\chi \gtrsim 1.667$.

\section{Method of the analysis} 
In this study, we estimate physical parameters of compact binaries, in which at least one component is a superspinar.
We use Bayesian inference to estimate physical parameters.
Bayes' theorem shows that a posterior probability density function $p(\boldsymbol{ \theta} |d)$ of parameters $\boldsymbol{ \theta}$ from observed data $d$ are given as
\be
p(\boldsymbol{ \theta} | d) = \f{\mathcal {L} (d| \boldsymbol{ \theta}) \, \pi (\boldsymbol{ \theta} ) }{Z},
\ee
where $\mathcal {L}(d| \boldsymbol{ \theta})$, $\pi(\boldsymbol{ \theta} )  $ and $Z$ are the likelihood function, a prior probability density function and evidence, respectively.
The likelihood for a single detector can be expressed as
\be
\mathcal {L}(d| \boldsymbol{ \theta}) \propto \exp\left[ -\f{1}{2} \left< \tilde{d  } (f) - \tilde{h } (f ;\boldsymbol{ \theta}) |  \tilde{d  } (f) - \tilde{h } (f ;\boldsymbol{ \theta}) \right>\right ], 
\label{like}
\ee
by assuming the detector noise is Gaussian, where  $\tilde{d  } (f)$ and  $ \tilde{h } (f)$ are Fourier transforms of the detected data and template waveform, respectively.
Here, 
\be
 \left <A  | B \right > \equiv 4 {\rm Re}\int ^ {f _{\rm high}}_{f _{\rm low}}  \f{A(f) B ^{*}( f)}{S_n(f)} df,
\label{match}
\ee
is the inner product weighted by the noise power spectral density (PSD) of the detector $S_n(f)$. 
The superscript $*$ shows the complex conjugate of the corresponding function.
The higher and lower cutoff frequencies of the data, $f _{\rm high}$ and $f _{\rm low}$, respectively, depend on the analysis.
Evidence is given as $Z = \int d\theta \, \mathcal {L}(d| \boldsymbol{ \theta}) \pi (\boldsymbol{ \theta} )$, which is used for model selection.
To obtain the posterior probability of parameters, we use \textsc{LALInference} \cite{Veitch:2014wba,lal}, which is one of the software suite of LIGO Algorithm Library.
Specifically, we use the nested sampling algorithm for stochastic samplings \cite{skilling, Veitch:2009hd}.

Since we do not know any physics at and after the merger of superspinar binaries, we focus on the inspiral part of the binary evolution.
We use the TaylorF2 waveform model as a template waveform.
We use the waveform model up to the 3.5 PN order for the phase and up to the 3 PN  order for the amplitude, where the point particle and the spin effects are included \cite{Dhurandhar:1992mw,Buonanno:2009zt,Blanchet:2013haa, Khan:2015jqa}.
Other waveform models that include merger and ringdown parts are calibrated by numerical waveforms, where black holes or neutron stars are assumed.
Therefore, although the TaylorF2 waveform model becomes inappropriate in the late inspiral phase \cite{Khan:2015jqa}, this waveform is the only choice to apply the spin larger than $|\chi| >1$.
Since the detectability of the spin value above the Kerr bound becomes worse for misaligned spin binaries compared to aligned spin ones for the same spin magnitude \cite{VanDenBroeck:2006ar}, restricting to aligned binaries may be a reasonable choice as a first step.
The improvement of waveform models valid for $|\chi| >1$ should be addressed as a future task.

We first perform injection studies, i.~e.~the data $\tilde {d} (f) $ in Eq.~\eqref{like} is replaced by a simulated signal.
Since our aim is to investigate the systematic bias due to the spin prior, we do not add simulated detector noise to the injection waveforms.
We inject the TaylorF2 waveform model  with $|\chi| >1 $ for either or both component spins.
Then, we estimate the binary parameters with two spin prior cases, $|\chi_{1,2}| \le 1$ and $|\chi_{1,2} | \le 1.5$, to see whether there is a bias in parameter estimation due to the limited spin prior range.
Here, we denote the former and latter priors as BH and SS priors, respectively.
Since the template waveform assumes that the component spins are aligned with the orbital angular momentum, we use the z-spin prior, which is equivalent to the prior that is uniform in magnitude and isotropic in orientation (see for details of the z-spin prior in Ref.~\cite{Lange:2018pyp}).

For injection waveforms, we consider two cases for the mass ratio, $q \equiv m_2/m_1=0.1$ and 0.5, where $m_1$ and $m_2$ are component masses with $m_1 > m_2$.
We fix the injection mass parameters as $(m_1, \, m_2) = (30 \,  M_{\odot}, \, 3 \, M_{\odot} )$ for $q=0.1$ and $(m_1, \,m_2) = (10 \, M_{\odot}, \, 5 \, M_{\odot})$ for $q=0.5 $.
For each case, we consider several injection spin values $\chi_{1,2} = (|0.1|, \, |0.5|, \, |0.8|, \, |1.1|)$, where at least one component of the binaries has $\chi=|1.1|$.
The luminosity distance $d_L$ and the inclination angle $\iota$ are set to 200 Mpc and 0 for injection, respectively.
The higher cutoff frequency  $f_{\rm high } $ is effectively  $f_{\rm ISCO }$ of the injection waveforms, which is $f_{\rm ISCO } = 133$ Hz for $q=0.1$ and $f_{\rm ISCO } = 293$ Hz for $q=0.5$.
The lower cutoff frequency is $f_{\rm low} = 20$ Hz.
We estimate all binary parameters: $(m_1, \, m_2, \, \chi_1, \,  \chi_2, \, d_L, \, \alpha, \, \delta,  \, \iota, \, \psi)$, where $\alpha $, $\delta$, and $\psi$ are the right ascension, declination and polarization angle, respectively.
We consider two Advanced LIGO detectors with their design sensitivity \cite{TheLIGOScientific:2014jea,ligopsd}  and one Advanced Virgo detector with its design sensitivity \cite{TheVirgo:2014hva}.

We next analyze the black hole binary merger events GW170608 and GW190814, which have a long inspiral phase with large signal to noise ratio among observed black hole binary events so far \cite{Abbott:2020jks}. 
We use the TaylorF2 waveform as a template with BH and SS priors.

There are two caveats to use the TaylorF2 waveform model.
Firstly, the TaylorF2 waveform terminates at $f_{\rm ISCO}$ of the corresponding total mass of the Schwarzschild black hole.
As explained in the previous section, $f_{\rm ISCO}$ depends on the spin, where the larger positive/negative spin gives larger/smaller values of $f_{\rm ISCO}$ compared to the Schwarzschild black hole case.
Therefore, we restrict ourselves to use the injection waveform for $\chi_1 > 0$, otherwise the injected waveform extends beyond the true $f_{\rm ISCO}$ value and the analysis may be inappropriate. 
Secondly, the abrupt cutoff of the waveform may cause a systematic bias in parameter estimation \cite{Mandel:2014tca}.
In our study, since we are interested in the bias due to the different spin prior range and the situation is the same for both spin priors, we do not further discuss about the bias from the abrupt cutoff of the waveform.
Although, we show some results with different $f_{\rm high}$ in Appendix A.

\section{Results} 

\subsection{Parameter estimation on injected waveforms} 
In this subsection, we show the marginalized posteriors of mass and spin parameters as well as the luminosity distance for some injection cases.
In all cases, the detector frame mass parameters are shown. 
\subsubsection{The case of mass ratio $q=0.1$} 

\begin{figure*}[t]
\includegraphics[scale=0.56]{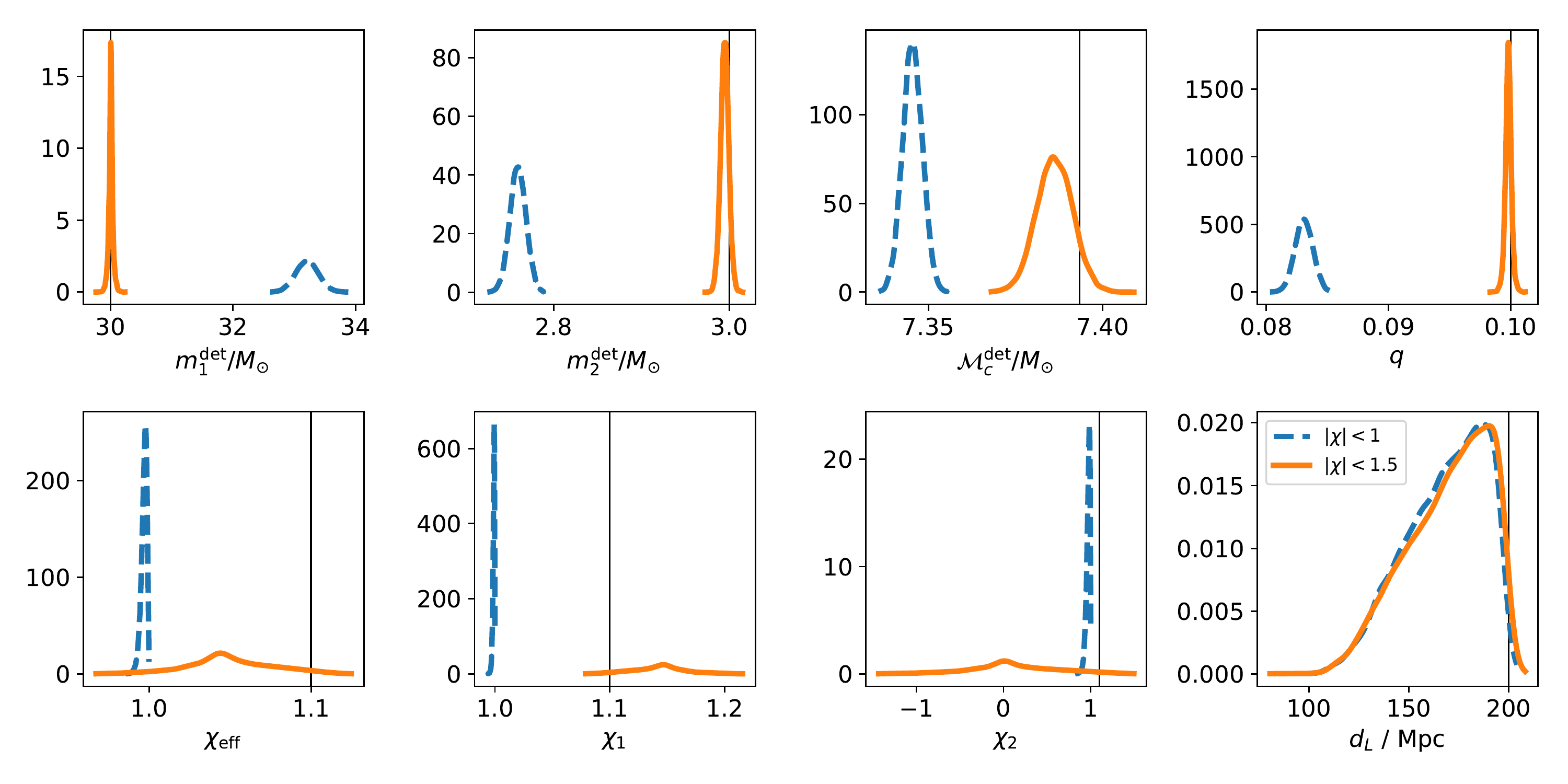}
\caption{Posterior distributions for mass parameters, spin parameters and the luminosity distance. Injected values are $(m_1, m_2, \chi_1, \chi_2) = (30 M_{\odot}, 3 M_{\odot}, 1.1, 1.1 )$. Blue and orange curves correspond to the cases when the spin prior is $|\chi_{1,2}| \le 1$ and $|\chi| \le 1.5$, respectively.  
The vertical lines correspond to the injected values.}
\label{kde-a1_1.1-a2_1.1_q0.1}
\end{figure*}

\begin{figure*}[t]
\includegraphics[scale=0.56]{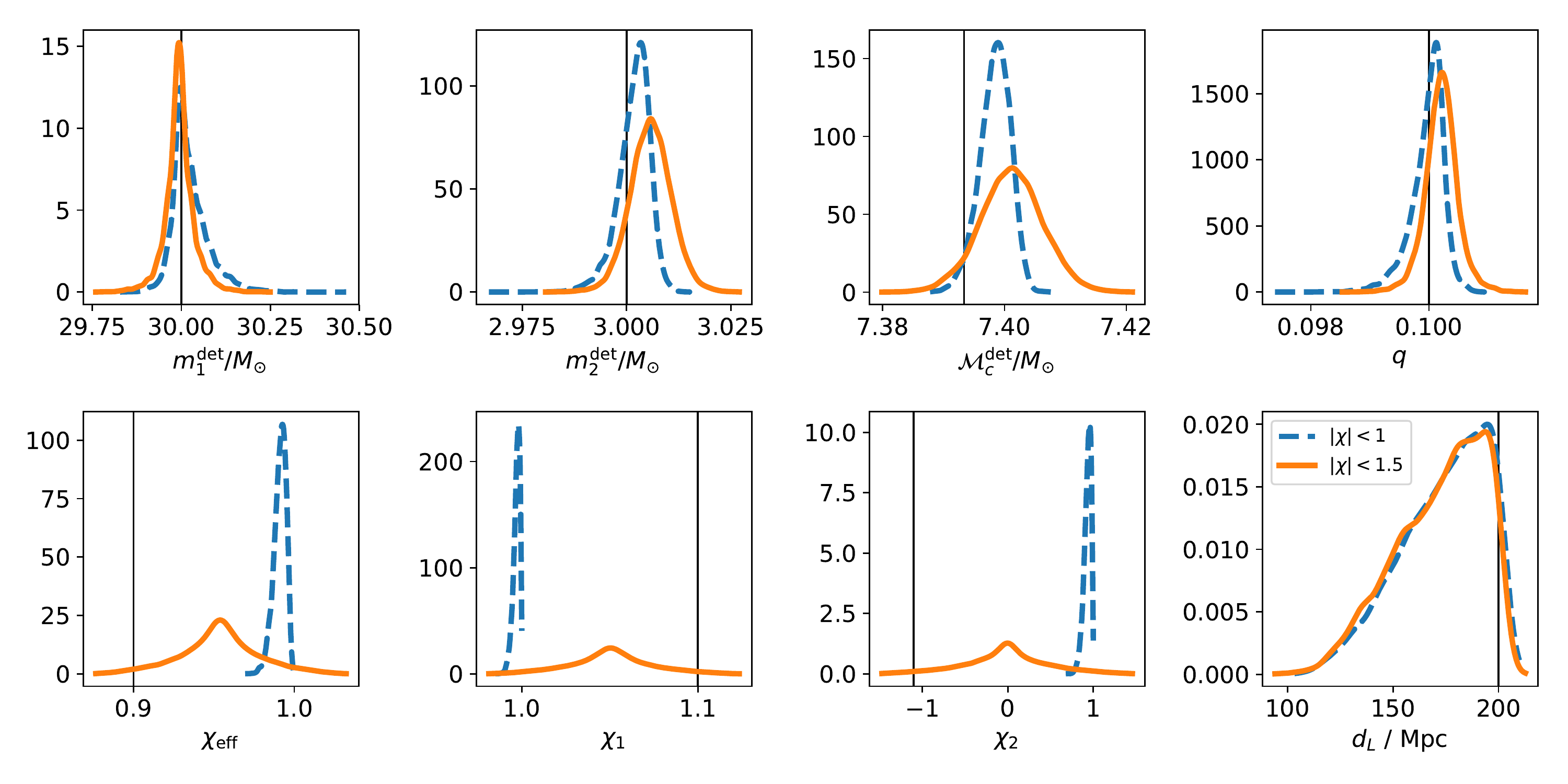}
\caption{Same as Fig.~\ref{kde-a1_1.1-a2_1.1_q0.1}, but for the injection with $(m_1, m_2, \chi_1, \chi_2) = (30 M_{\odot}, 3 M_{\odot}, 1.1, -1.1 )$. }
\label{kde-a1_1.1-a2_m1.1_q0.1}
\end{figure*}

\begin{figure*}[t]
\includegraphics[scale=0.56]{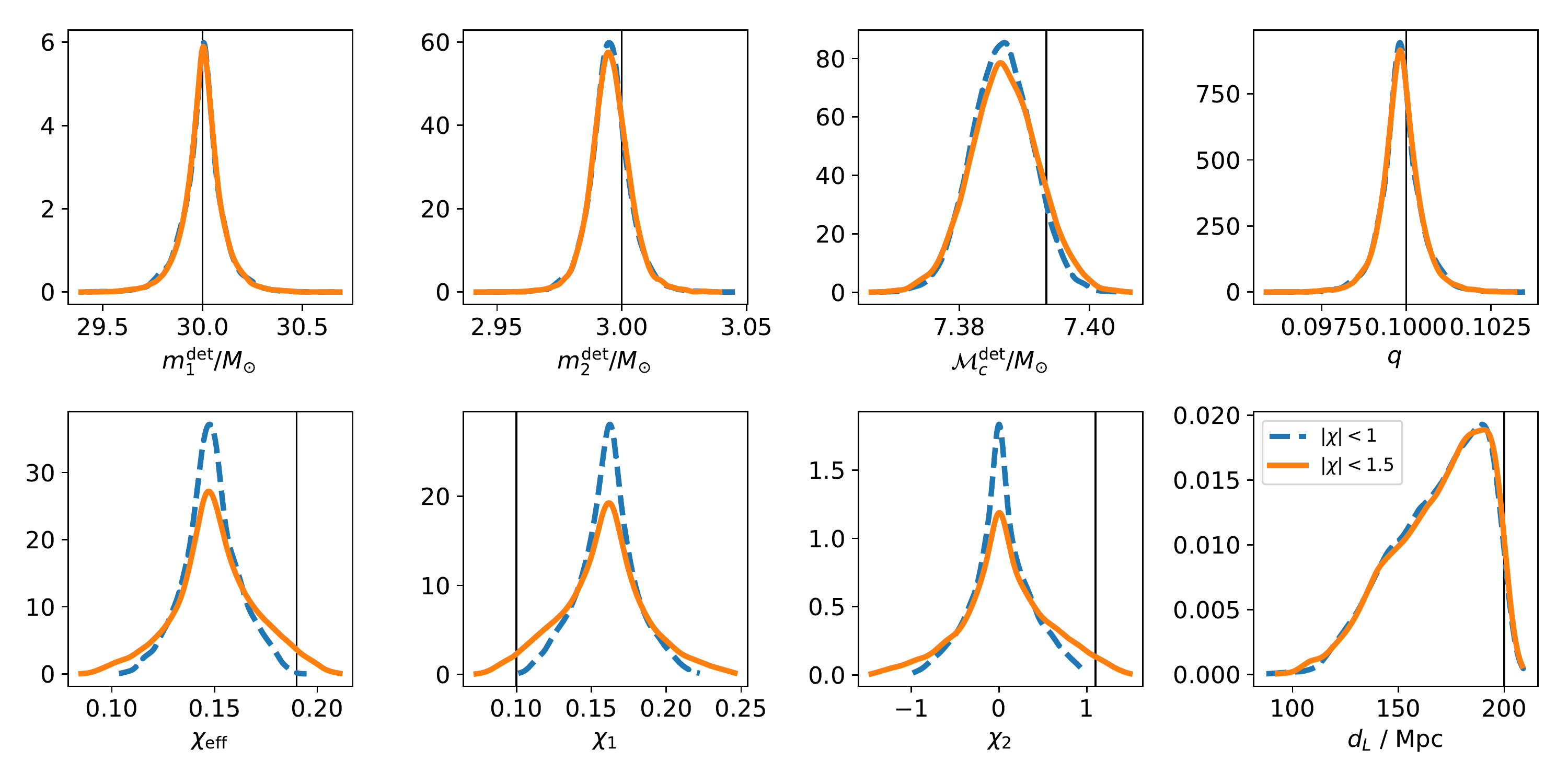}
\caption{Same as Fig.~\ref{kde-a1_1.1-a2_1.1_q0.1}, but for the injection with $(m_1, m_2, \chi_1, \chi_2) = (30 M_{\odot}, 3 M_{\odot}, 0.1, 1.1 )$.}
\label{kde-a1_0.1-a2_1.1_q0.1}
\end{figure*}

\begin{figure*}[t]
\includegraphics[scale=0.56]{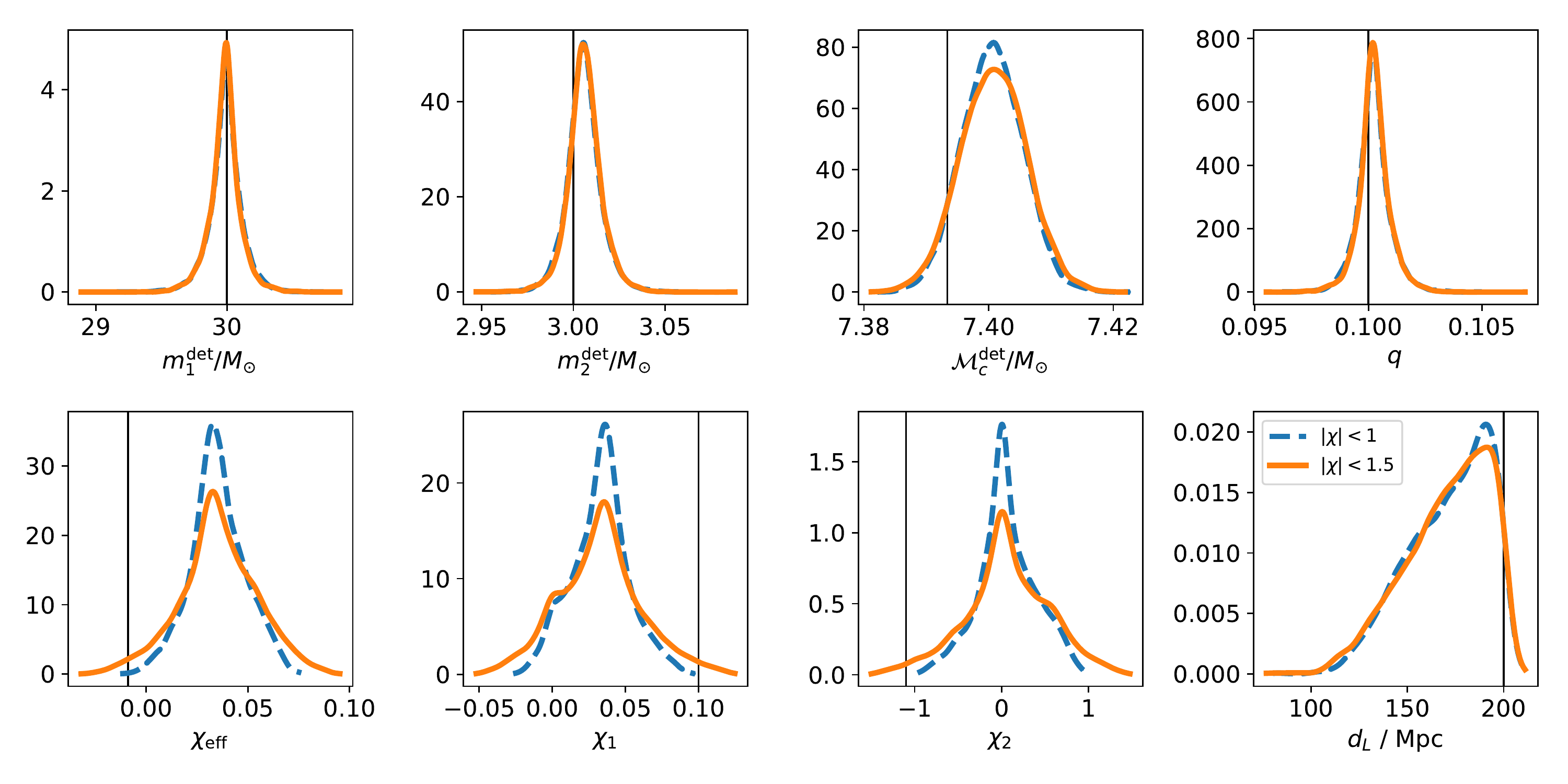}
\caption{Same as Fig.~\ref{kde-a1_1.1-a2_1.1_q0.1}, but for the injection with $(m_1, m_2, \chi_1, \chi_2) = (30 M_{\odot}, 3 M_{\odot}, 0.1, -1.1 )$. }
\label{kde-a1_0.1-a2_m1.1_q0.1}
\end{figure*}

We first show the posterior distributions of mass parameters ($m_1,m_2,q, \mathcal{M}_c$), spin parameters ($\chi_{\rm eff}, \chi_1, \chi_2$) and the luminosity distance $d_L$ for the mass ratio $q=0.1$ case in Figs.~\ref{kde-a1_1.1-a2_1.1_q0.1} -- \ref{kde-a1_0.1-a2_m1.1_q0.1}.
Here, the chirp mass $ \mathcal{M}_c = (m_1 m_2) ^{3/5} /(m_1 + m_2) ^{1/5}$ and the effective inspiral spin $\chi_{\rm eff} = (m_1 \chi_1 + m_2 \chi_2) / ( m_1 + m_2)$ are the most well determined mass and spin parameters in the inspiral waveform \cite{Ajith:2009bn}.
Figure \ref{kde-a1_1.1-a2_1.1_q0.1} shows the results when the waveform with $(m_1, m_2, \chi_1, \chi_2) = (30 M_{\odot}, 3 M_{\odot}, 1.1, 1.1 )$ is injected, which gives the most significant bias in parameter estimation due to the different spin prior range.
Naturally, the spin parameters are not correctly estimated for $|\chi_{1,2}| \le 1$ prior, since the injected value is outside the prior range.
Posteriors of all spin parameters concentrate on the positive prior bound for this prior.
Moreover, not only the spin parameters, the mass parameters are also biased by the limited spin prior range.
When the spin prior is extended to $|\chi _{1,2}| \le 1.5$, the mass parameters are well estimated.
While the posterior distributions of $\chi_{\rm eff}$ and $\chi_1$ are constrained around the injected values, $\chi_2$ becomes undetermined.
This is natural because $\chi_{\rm eff}$ is dominated by $\chi_1$ for the asymmetric mass ratio.
We see the similar tendency for different $\chi_2$ as shown in Fig.~\ref{kde-a1_1.1-a2_m1.1_q0.1}, which shows the posteriors of $(\chi_1,\chi_2) =(1.1,-1.1)$ injection.
The bias in mass parameters becomes smaller as $\chi_2$ becomes smaller. 
Although $\chi_2 = -1.1$ is injected, the posterior concentrates on the positive bound for the prior $|\chi_{1,2}| \le 1$.

Figures \ref{kde-a1_0.1-a2_1.1_q0.1} and \ref{kde-a1_0.1-a2_m1.1_q0.1} are the same as Fig.~\ref{kde-a1_1.1-a2_1.1_q0.1}, but the injected spins are $(\chi_1,\chi_2) =(0.1,1.1)$ and $(\chi_1,\chi_2) =(0.1,-1.1)$, respectively.
In these cases, we do not see much differences in estimation of mass and spin parameters due to the different spin prior range unlike the $\chi_1 = 1.1$ injection cases.
For both spin prior cases, the posteriors of mass parameters include the injected value, on the other hand, this is not the case for spin parameters; the injected value is included only for the $|\chi_{1,2}| \le 1.5$ prior case.
From Figs.~\ref{kde-a1_1.1-a2_1.1_q0.1} -- \ref{kde-a1_0.1-a2_m1.1_q0.1}, we can see that $\chi_{\rm eff}$ tends to underestimate (overestimate) the injected value for $\chi_2 > 0$ ($\chi_2 <0$).

We do not see much differences in the matched filter signal-to-noise ratio (SNR) between two spin priors.
For example, the SNR is $\sim 80$ and $\sim 78$ for $|\chi_{1,2}| \le 1.5$ prior and $|\chi_{1,2}| \le 1$ prior, respectively, for $(\chi_1,\chi_2) =(1.1,1.1)$ injection. 

The 90\% symmetric credible regions of spin parameters for the $|\chi _{1,2}| \le 1.5$ prior together with values that give the maximum likelihood are summarized in Table \ref{credible}.
We can see that $\chi_{\rm eff}$ and $\chi_1$ show good estimates for the both aligned case ($\chi_{1,2} > 0$), while the estimate of $\chi_2$ is poor and the error is large for any injected value, as expected for asymmetric mass ratio binaries.

\subsubsection{The case of mass ratio $q=0.5$} 

\begin{figure*}[t]
\includegraphics[scale=0.56]{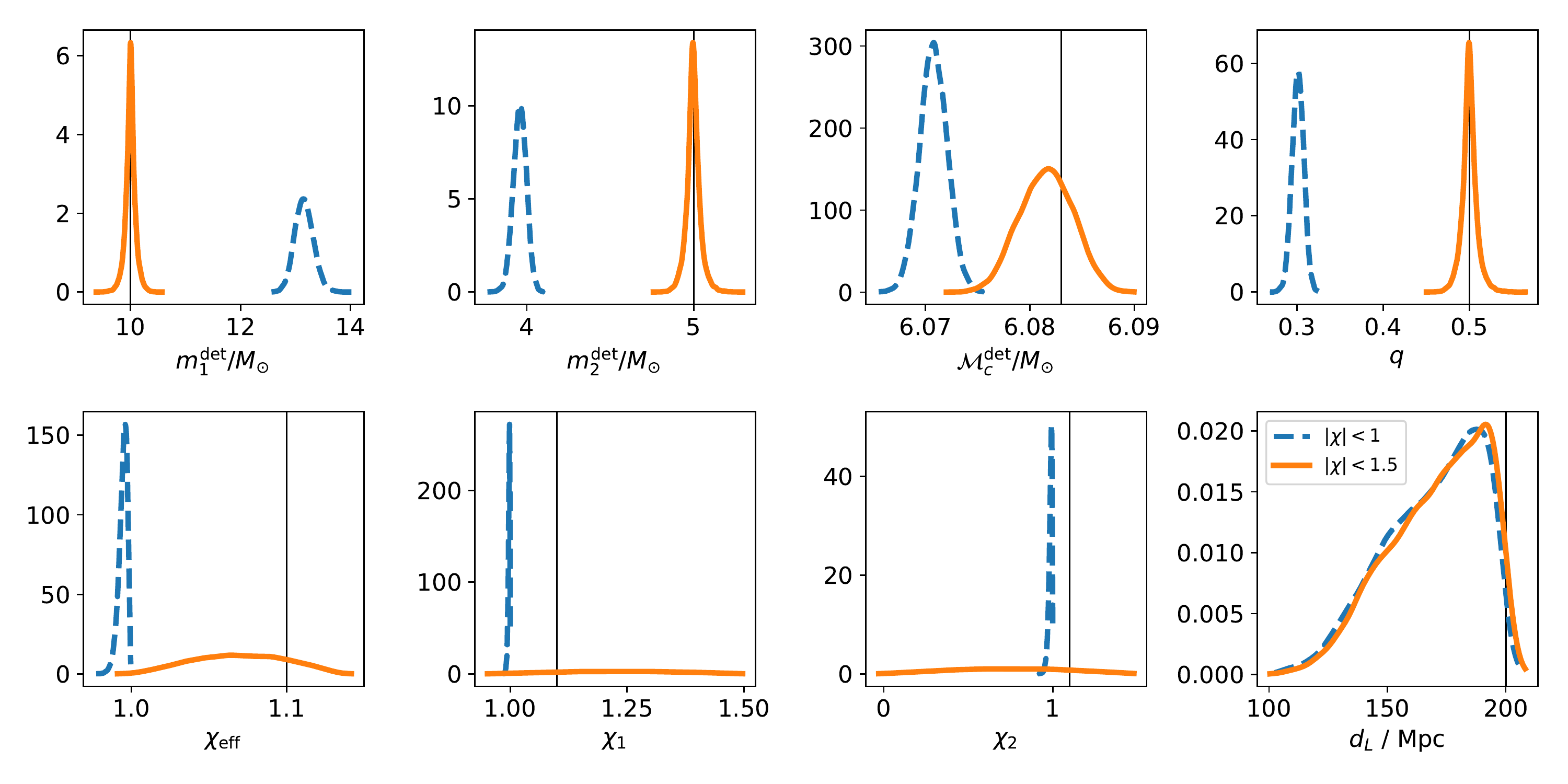}
\caption{Same as Fig.~\ref{kde-a1_1.1-a2_1.1_q0.1}, but for $(m_1, m_2, \chi_1, \chi_2) = (10 M_{\odot}, 5 M_{\odot}, 1.1, 1.1 )$. }
\label{kde-a1_1.1-a2_1.1_q0.5}
\end{figure*}

\begin{figure*}[t]
\includegraphics[scale=0.56]{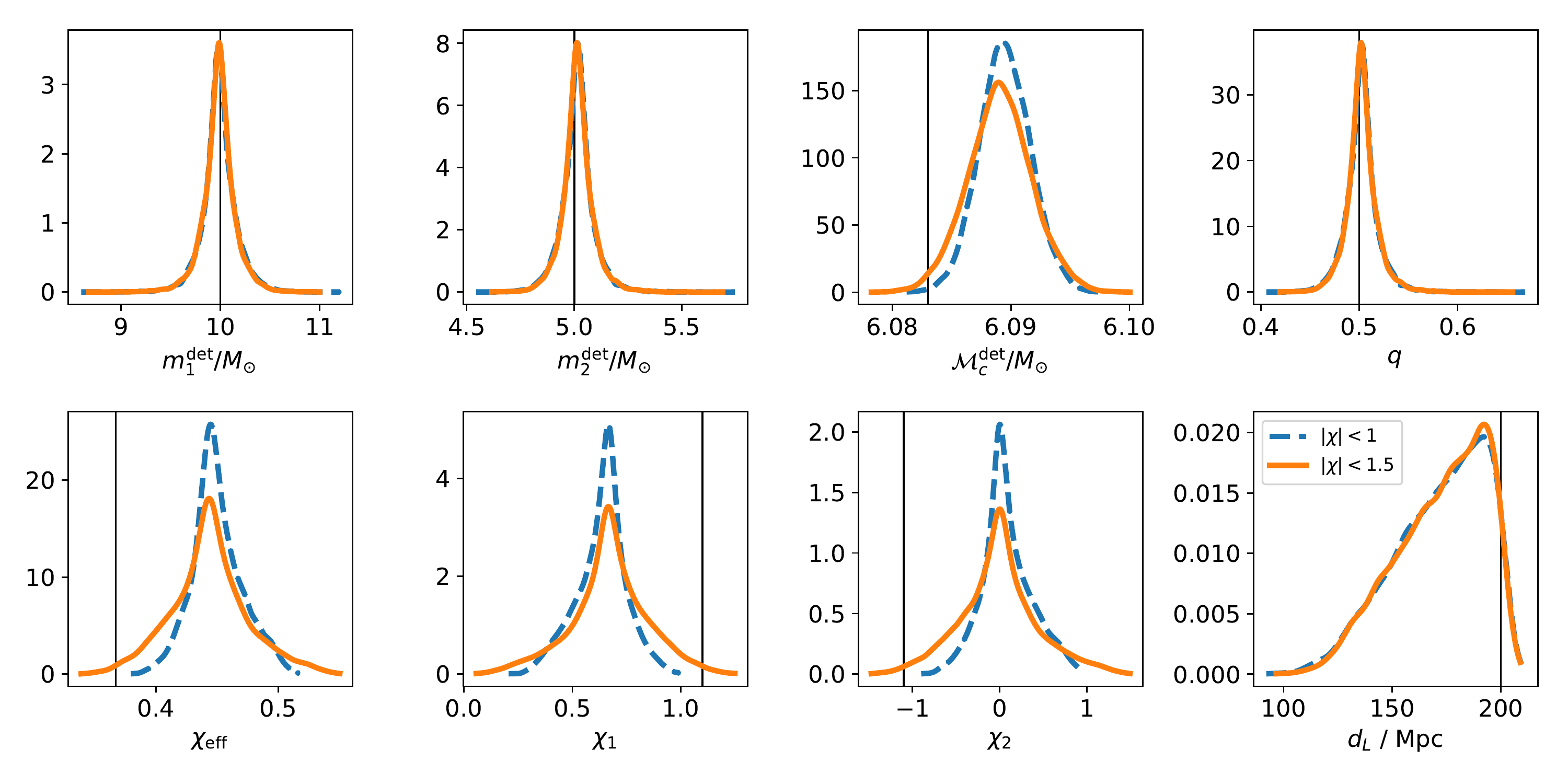}
\caption{Same as Fig.~\ref{kde-a1_1.1-a2_1.1_q0.1}, but for $(m_1, m_2, \chi_1, \chi_2) = (10 M_{\odot}, 5 M_{\odot}, 1.1, -1.1 )$. }
\label{kde-a1_1.1-a2_m1.1_q0.5}
\end{figure*}

\begin{figure*}[t]
\includegraphics[scale=0.56]{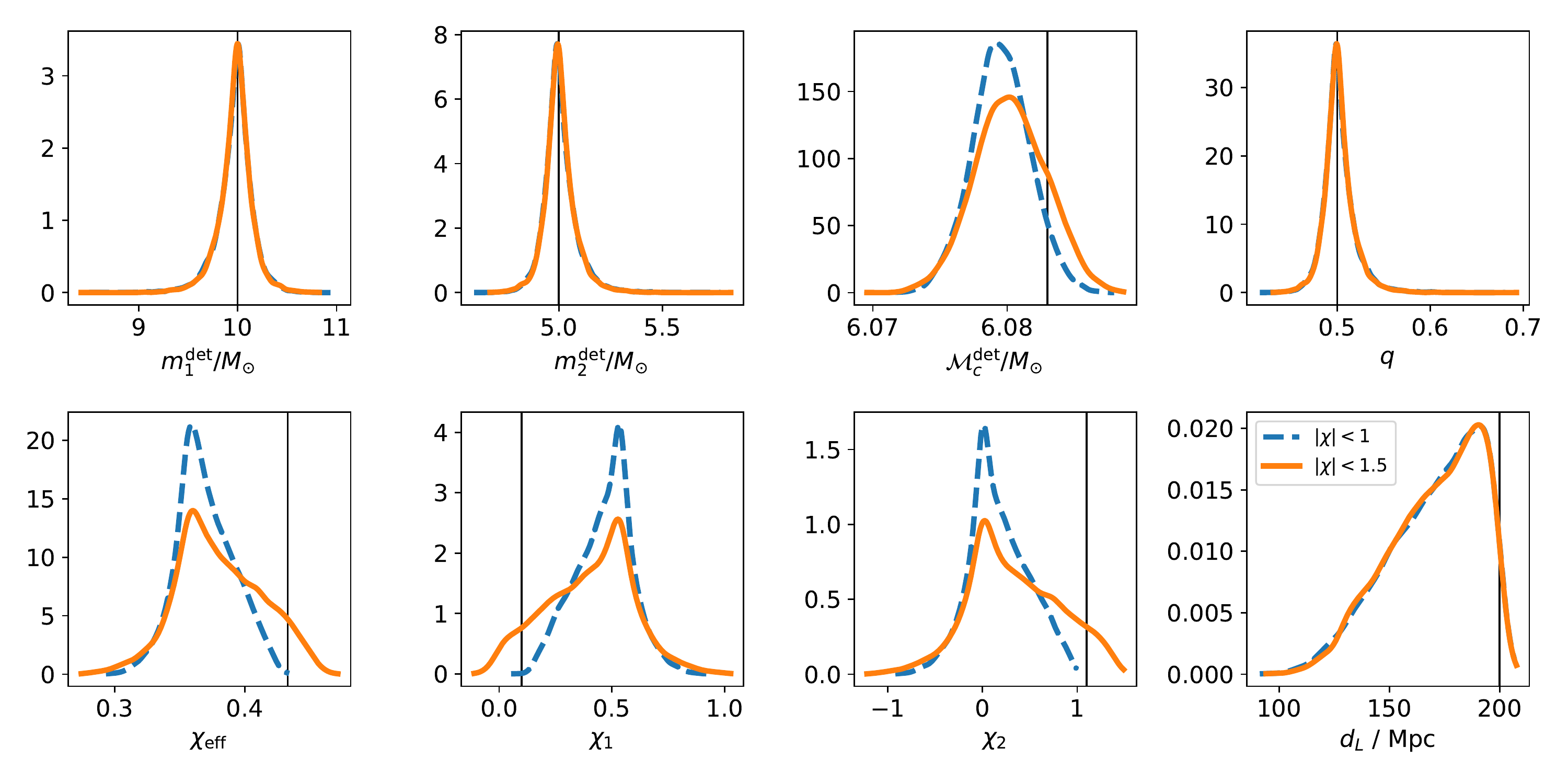}
\caption{Same as Fig.~\ref{kde-a1_1.1-a2_1.1_q0.1}, but for the injection with $(m_1, m_2, \chi_1, \chi_2) = (10 M_{\odot}, 5 M_{\odot}, 0.1, 1.1 )$.}
\label{kde-a1_0.1-a2_1.1_q0.5}
\end{figure*}

\begin{figure*}[t]
\includegraphics[scale=0.56]{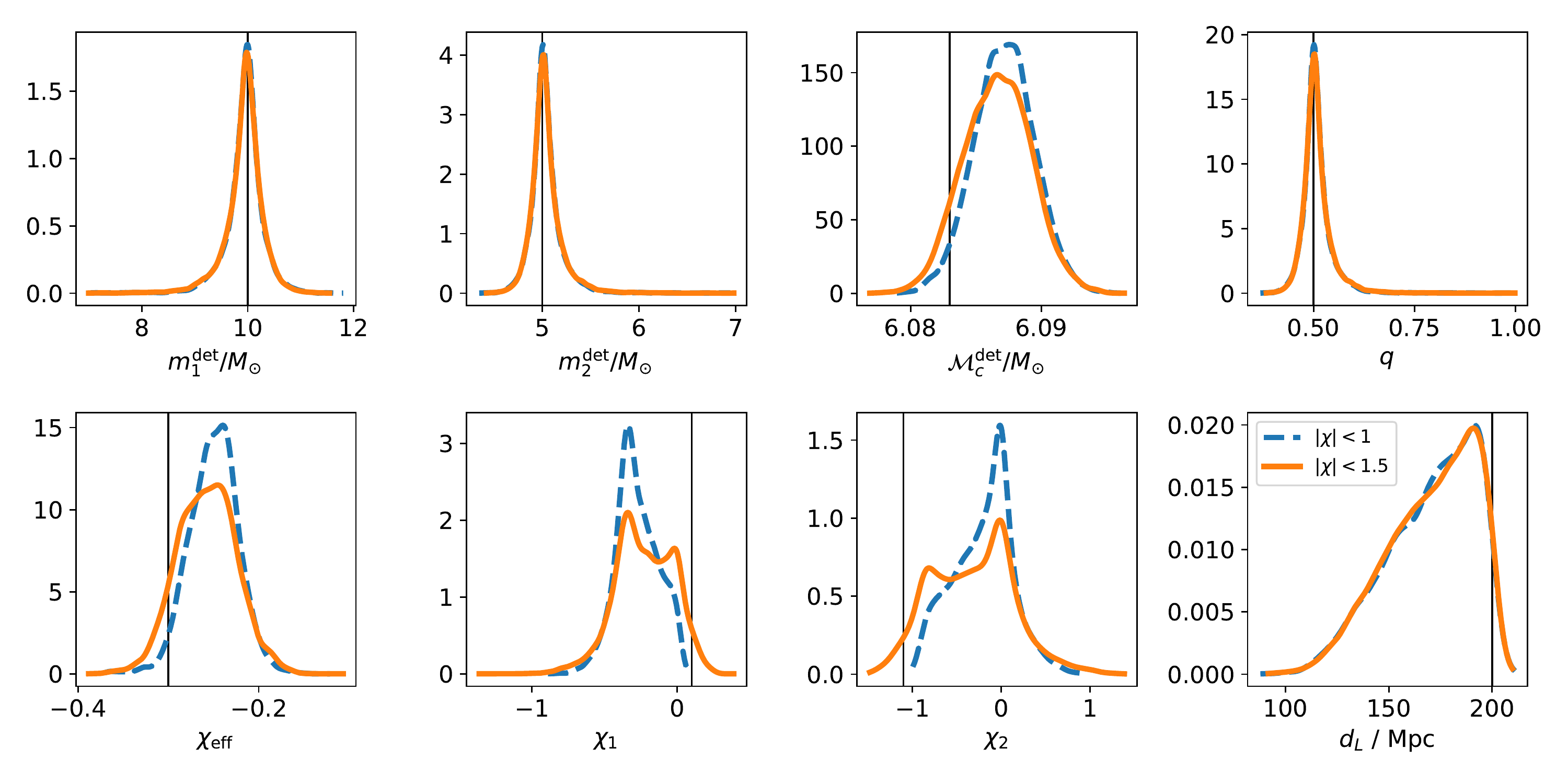}
\caption{Same as Fig.~\ref{kde-a1_1.1-a2_1.1_q0.1}, but for the injection with $(m_1, m_2, \chi_1, \chi_2) = (10 M_{\odot}, 5 M_{\odot}, 0.1, -1.1 )$.}
\label{kde-a1_0.1-a2_m1.1_q0.5}
\end{figure*}

Next, we show the results for the mass ratio $q=0.5$ [$(m_1, m_2) = (10 M_{\odot}, 5 M_{\odot}) $] in Figs.~\ref{kde-a1_1.1-a2_1.1_q0.5} -- \ref{kde-a1_0.1-a2_m1.1_q0.5}.
Figure \ref{kde-a1_1.1-a2_1.1_q0.5} shows the posteriors when $(\chi_1, \chi_2) =(1.1, 1.1)$ are injected.
Similar to Fig.~\ref{kde-a1_1.1-a2_1.1_q0.1}, we see biases in estimation of mass and spin parameters due to the spin prior range.
Posteriors of all spin parameters concentrate on the positive prior bound for $|\chi_{1,2}| \le 1$ prior.
For $q=0.1$, $\chi_2$ is undetermined for $|\chi_{1,2}| \le 1.5$ prior, while it is constrained to be positive for $q=0.5$.
From Fig.~\ref{kde-a1_1.1-a2_m1.1_q0.5}, which shows the posteriors when $(\chi_1, \chi_2) =(1.1, -1.1)$ are injected, there is almost no bias in estimation of mass and spin parameters.
The secondary spin $\chi_2 $ is undetermined for both spin prior cases.
The posteriors obtained from $(\chi_1, \chi_2) =(0.1, 1.1)$ and $(\chi_1, \chi_2) =(0.1, -1.1)$ injections are shown in Figs.~\ref{kde-a1_0.1-a2_1.1_q0.5} and \ref{kde-a1_0.1-a2_m1.1_q0.5}, respectively.
Again we do not see much biases in estimation of mass and spin parameters due to the spin prior range from these figures.
The secondary spin is undetermined for both priors for these cases.
For $\chi_{\rm eff}>0$ injection, $\chi_{\rm eff}$ and $\chi_1$ are estimated as positive values, which can be seen in Figs.~\ref{kde-a1_1.1-a2_1.1_q0.5} --  \ref{kde-a1_0.1-a2_1.1_q0.5}.
On the other hand, for $\chi_{\rm eff}<0$ injection, the posterior of $\chi_1$ mainly distributed in negative values even a positive value is injected as shown in Fig.~\ref{kde-a1_0.1-a2_m1.1_q0.5}.
Similar to the $q=0.1$ case, $\chi_{\rm eff}$ tends to underestimate (overestimate) the injected value for $\chi_2 > 0$ ($\chi_2 < 0 $).

Like the $q=0.1$ case, we do not see much difference in SNR between two spin priors.
Even for the most significant bias case, $(\chi_1, \chi_2) =(1.1, 1.1)$ injection, SNR is $\sim 74 $ and $\sim 73 $ for $|\chi_{1,2}| \le 1.5$ prior and $|\chi_{1,2}| \le 1$ prior, respectively.

The 90\% symmetric credible regions of spin parameters for the $|\chi_{1,2}| \le 1.5$ prior and values of those parameters that give the maximum likelihood are also summarized in Table \ref{credible}.
We can see that the 90\% regions of all spin parameters include the injected values for the both aligned case. 
A small bias in the estimations for the $(\chi_1,\chi_2)=(1.1,-1.1)$ injection case might be caused by the prior since the values at the maximum likelihood are close to the injected values.
A bias also seen in the estimations for the $(\chi_1,\chi_2) = (0.1,-1.1)$ injection case might be caused by the abrupt cutoff of the TaylorF2 waveform.
The values at the maximum likelihood deviate from the injected ones in this case.

\begin{center}
\begin{table*}[htbp]
 \caption{The symmetric 90\% credible regions of $(\chi_{\rm eff} , \chi_1, \chi_2)$ for the spin prior $|\chi_{1,2}| \le 1.5$. Values of spin parameters that give the maximum likelihood is also shown.}
\begin{tabular}{l l l l l l l l l l }
\hline \hline
\multicolumn{3}{c}{Injected parameters}  &\multicolumn{3}{c}{90\% symmetric credible regions} &\multicolumn{3}{c}{Maximum likelihood} \\
  $(m_1,m_2)$ & $\chi_{\rm eff}$&$( \chi_1, \chi_2)$&$\chi_{\rm eff}$ & $\chi_1$ & $\chi_2$ &$\chi_{\rm eff}$ & $\chi_1$ & $\chi_2$\\ \hline
$(30 M_{\odot},3M_{\odot})$ & 1.1&(1.1,1.1)& $(1.01,1.09)$ & $(1.10, 1.18)$ & $(-0.72,1.00)$ & 1.09& 1.11&0.85\\ 
		& 0.9 &$(1.1,-1.1)$&  $(0.91,0.99)$ & $(1.01,1.09) $ &$(-0.86,0.82)$ & 0.95 & 1.06 & $-0.18$ \\
		&0.19 &(0.1,1.1) &  $(0.12,0.19)$ & $(0.11,0.21) $ &$(-0.77,0.95)$ & 0.14& 0.17& $-0.15$ \\ 
		& $-0.01 $ & $(0.1,-1.1 )$& (0.00, 0.07)& $(-0.01,0.08)$ & $(-0.80,0.87)$ &0.02 &0.05 &$-0.25$\\ 
 $(10M_{\odot},5M_{\odot})$ & 1.1&$(1.1,1.1)$& $(1.02,1.12) $& $(1.03,1.43)$ & $(0.2,1.28)$ &1.12 & 1.03 &$1.28$\\ 
		&0.37 &$(1.1,-1.1)$&  $(0.39,0.50)$& $(0.35,0.97) $ &$(-0.76,0.79)$  &0.38 & 1.03 & $-0.91$\\
		&0.43&(0.1,1.1) & $ (0.33,0.44)$& $ (0.06,0.69)$ &$ (-0.39,1.18)$  &0.42 & 0.16 & 0.94 \\ 
		&$-0.3$ & $ (0.1,-1.1 )$& $(-0.31,-0.21)$& $(-0.55,0.08)$ & $(-1.04,0.43)$ & $-0.25$ &$-0.29$ & $ -0.15$\\ 
  \hline \hline
\end{tabular}
  \label{credible}
\end{table*}
\end{center}

In summary for this subsection, estimation of spin and mass parameters are biased for $|\chi_{1,2}| \le 1$ prior when $\chi_1 = \chi_2= 1.1 $ are injected.
In this case, posteriors of all spin parameters concentrate on the positive prior bound for $|\chi_{1,2}| \le 1$ prior.
This suggests that such behavior of the spin posteriors can be the first clue of the detection of superspinars.
On the other hand, the bias due to $|\chi_{1,2}| \le 1$ prior becomes smaller for small $\chi_{\rm eff}$ injections.
We will discuss the bias in estimated mass and spin parameters in the next subsection.
We also observe that SNR does not show much difference with respect to the different spin prior range, that is, SNR cannot be a good indicator to deicide which prior range is appropriate.
This topic will be addressed in Sec.~\ref{model}.

\subsection{Discussions of biases in mass and spin parameters} 
\begin{figure}[htbp]
\includegraphics[scale=0.47]{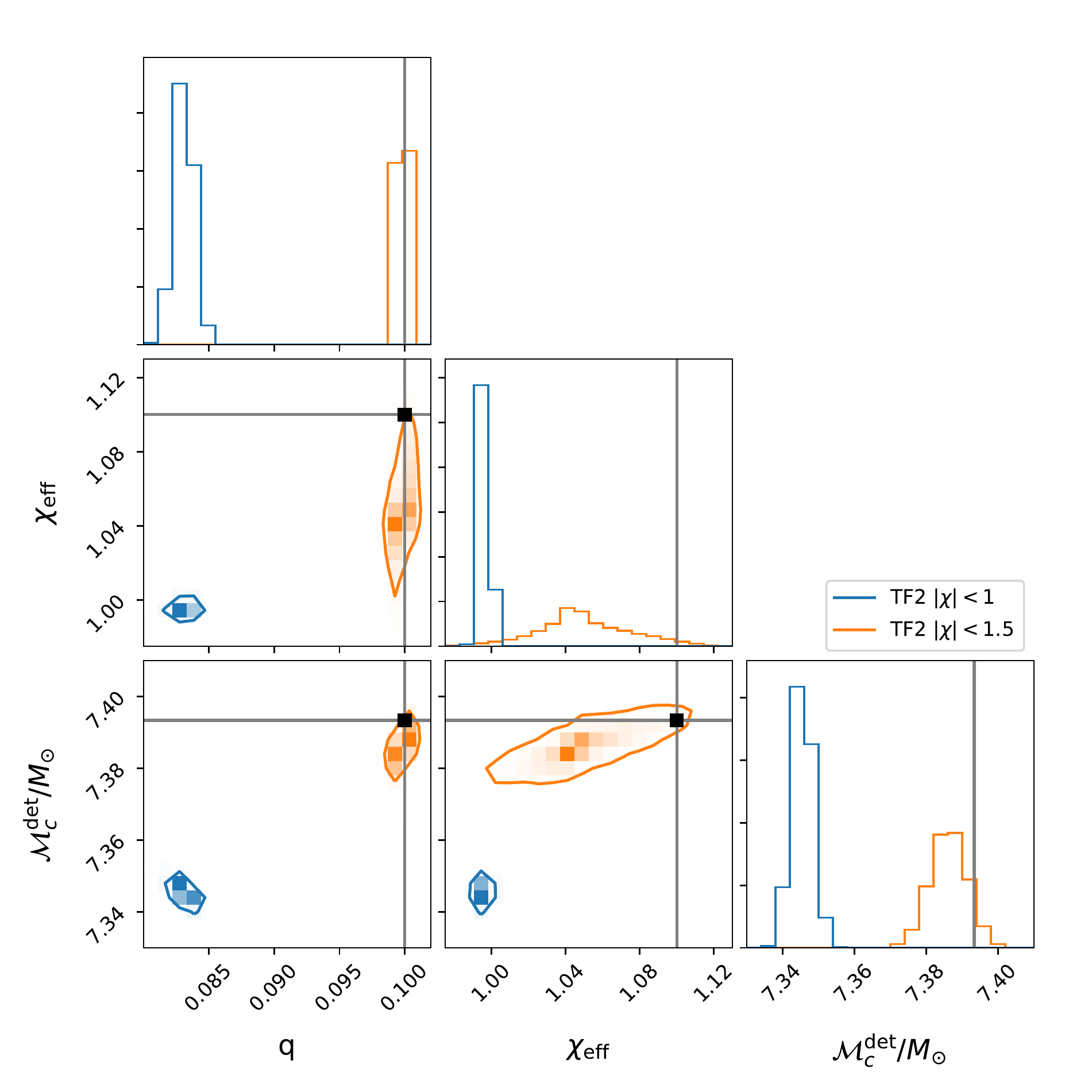}
\caption{Comparison of two-dimensional posterior distributions of $(q, \chi_{\rm eff}, \mathcal{M}_c^{\rm det} )$ for the injected values $( m_1, m_2, \chi_1, \chi_2 ) =(30 M_{\odot}, 3 M_{\odot}, 1.1, 1.1)$.
Blue and orange contours show 90\% credible regions of corresponding parameters for BH and SS priors, respectively.
Injected values are shown by black lines and squares.}
\label{corner}
\end{figure}

In this subsection, we discuss the biases in estimated mass and spin parameters caused by the limited spin prior range.
First, we consider whether the bias in mass and spin parameters for $\chi_1 = \chi_2 = 1.1$ injection, as shown in Fig.~\ref{kde-a1_1.1-a2_1.1_q0.1}, can be explained by a degeneracy between mass and spin parameters.
Several studies show that there is a degeneracy between the mass ratio and the effective spin for the inspiral dominant waveform \cite{Purrer:2015nkh,Chatziioannou:2018wqx,170817, gwtc-1}.
In Fig.~\ref{corner}, we compare two-dimensional posterior distributions of $(q, \chi_{\rm eff}, \mathcal{M}_c^{\rm det} )$ for the injected values $( m_1, m_2, \chi_1, \chi_2 ) =(30 M_{\odot}, 3 M_{\odot}, 1.1, 1.1)$. 
Although our study only use the inspiral waveform, we do not see a degeneracy between $q$ and $\chi_{\rm eff}$ in both prior cases.
For the $|\chi| \le 1.5$ prior case, we see a a positive correlation between $\mathcal{M} _{c}^{\rm det}$ and $\chi_{\rm eff}$.
However, it seems to be difficult to explain the bias in the results for the $|\chi| \le 1$ prior case from the correlation in  $\mathcal{M} _{c}^{\rm det}$ and $\chi_{\rm eff}$ for $|\chi| \le 1.5$.

\begin{figure}[htbp]
\includegraphics[scale=0.47]{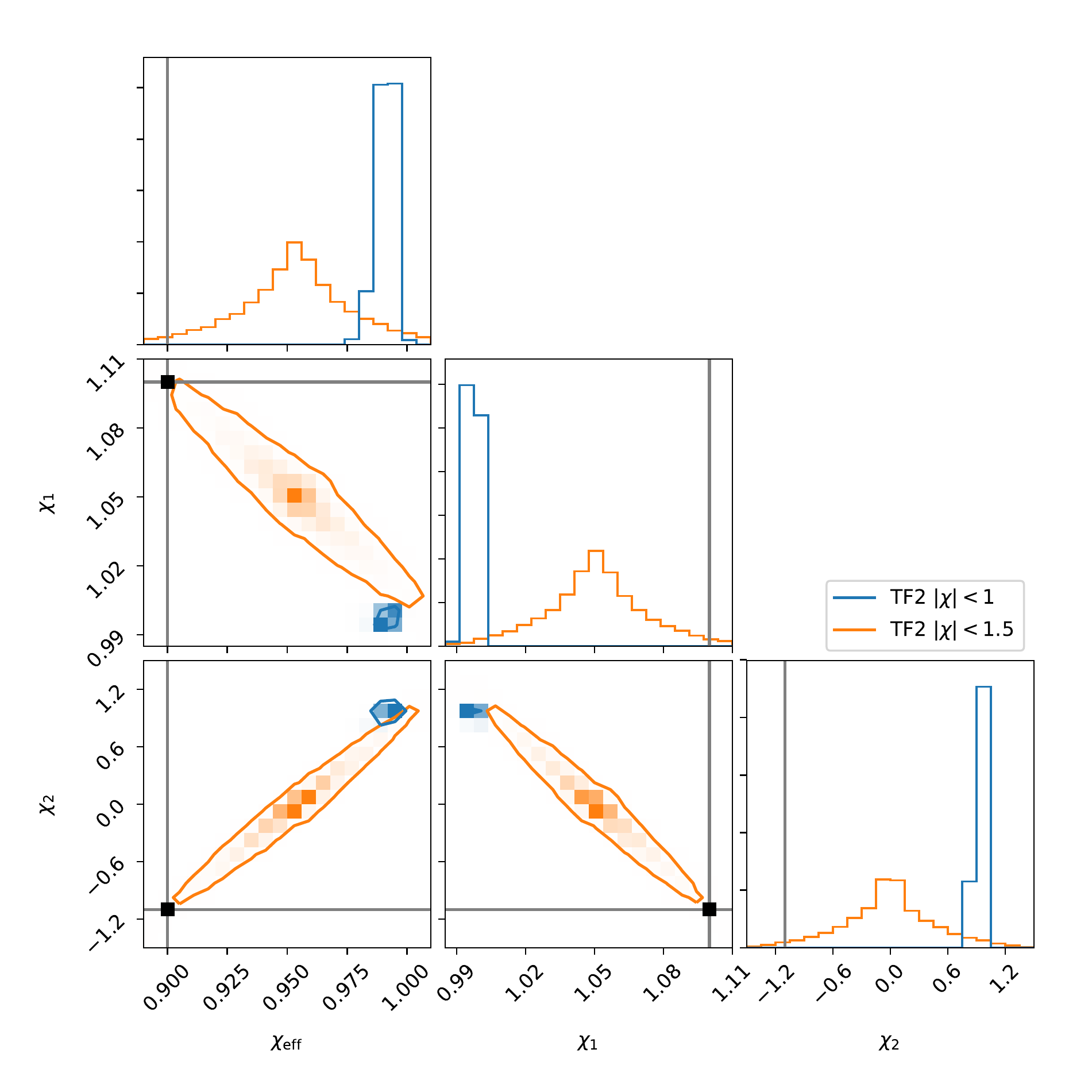}
\caption{Comparison of two-dimensional posterior distributions of $(\chi_{\rm eff}, \chi_1, \chi_2 )$ for the injected values $( m_1, m_2, \chi_1, \chi_2 ) =(30 M_{\odot}, 3 M_{\odot}, 1.1, -1.1)$.
Blue and orange contours show 90\% credible regions of corresponding parameters for BH and SS priors, respectively.
Injected values are shown by black lines and squares.}
\label{corner2}
\end{figure}

Next, we discuss the bias in estimated spin parameters for  $( m_1, m_2, \chi_1, \chi_2 ) =(30 M_{\odot}, 3 M_{\odot}, 1.1, -1.1)$ injection as shown in Fig.~\ref{kde-a1_1.1-a2_m1.1_q0.1}.
In this case, $\chi_2 $ is estimated as $\chi_2 \approx 1$ for the $|\chi| \le 1$ prior, even though $\chi_2=-1.1$ is injected.
Figure \ref{corner2} shows the two-dimensional posterior distributions of three spin parameters $(\chi_{\rm eff}, \chi_1, \chi_2 )$ for $( m_1, m_2, \chi_1, \chi_2 ) =(30 M_{\odot}, 3 M_{\odot}, 1.1, -1.1)$ injection.
For the $|\chi| \le 1.5$ prior, a negative correlation can be seen between $\chi_1$ and $\chi_2$.
Due to this correlation, when $\chi_1$ is restricted to $\chi_1 \le 1$, the allowed region for $\chi_2$ becomes $\chi_2 \approx 1$ even though the negative $\chi_2 $ is injected.

\subsection{Model selection for the injection studies} 
\label{model}
\begin{figure*}[t]
\includegraphics[scale=0.5]{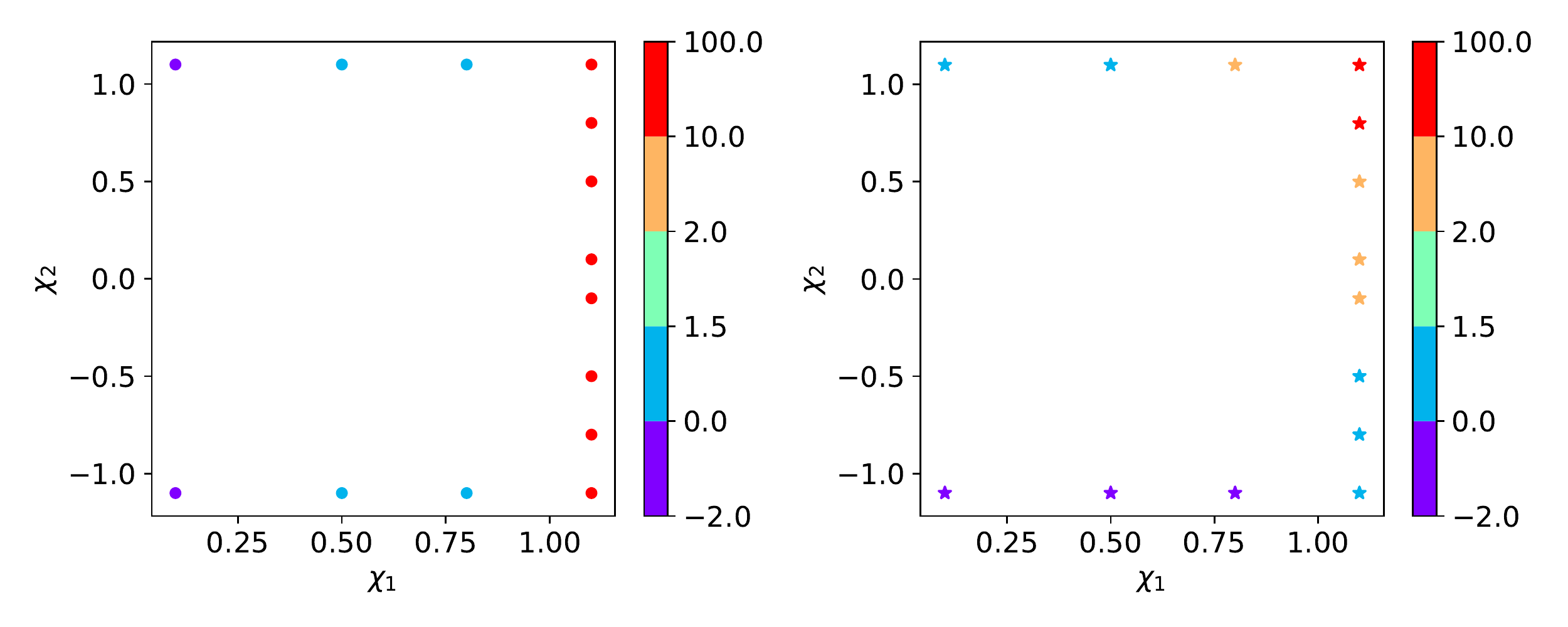}
\caption{$\log_{10}$ Bayes factor  against $\chi_1$ and $\chi_2$. Color bar shows the value of $\log_{10} (\rm BF ^{\rm SS}_{\rm BH})$, where SS denotes the case of spin prior ($|\chi| \le 1.5$) and BH denotes the spin prior ($|\chi| \le 1$).
Left and right panels show the cases when $(m_1, m_2) = (30 M_{\odot}, 3 M_{\odot} )$ and $(m_1, m_2) = (10 M_{\odot}, 5 M_{\odot} )$, respectively.
}
\label{BF}
\end{figure*}

In this subsection, we evaluate which spin prior is favored by model selection, in which the evidence from each prior is compared.
The ratio of the evidences is called Bayes factor.
Here, we compare the evidences for two prior assumptions SS and BH, where the Bayes factor is expressed as ${\rm BF^{\rm SS}_{\rm BH}} = Z^{\rm SS} /Z^{\rm BH}$.
Logarithmic Bayes factor of 2 is a threshold to claim the decisive evidence of the hypothesis of the numerator \cite{jeffreys}. 
We show  $\log_{10} {\rm BF^{\rm SS}_{\rm BH} }  $ for each injection in Fig.~\ref{BF}, where left and right panels show the injections for $q=0.1$ and $0.5$, respectively.
From the left panel of the figure, the SS assumption is strongly favored for $\chi_1 = 1.1$ injections, while no strong preference of the SS assumption for $\chi_1 = 0.1$ injections is shown although $\chi_2$ is a superspinar.
This means that we can claim the existence of a superspinar if the primary is a superspinar, while it is difficult to claim so if just the secondary is a superspinar.
The minimum value of $\log_{10} {\rm BF^{\rm SS}_{\rm BH}} $ is  $-0.323$ when $(\chi_1, \chi_2) = (0.1,-1.1)$ are injected.
From the right panel of the figure, for $q=0.5$, the SS assumption is strongly favored for $\chi_1=1.1$ and $ \chi_2 >0$ injections.
When $\chi_2 = -1.1$ is injected, there is no preference for the SS assumption if $\chi_1 <1.0$ is injected.
The result implies that for the moderate asymmetric mass ratio case, strong evidence of the SS assumption can be obtained when $\chi_1=1.1$ and $ \chi_2 > 0$ are injected.
For $q=0.5$, the minimum value of $\log_{10} {\rm BF^{\rm SS}_{\rm BH}} $ is  $-0.685$  when $(\chi_1, \chi_2) = (0.8,-1.1)$ are injected.
We should note that the larger prior range or volume tends to give smaller evidence because of an Occam factor \cite{Zevin:2020gxf}.

\subsection{Parameter estimation on black hole binary  events} 
We finally analyze two black hole binary events GW170608 and GW190814 with the extended spin prior using TaylorF2 waveform. 
The inspiral SNRs are $\sim 15$ and $\sim 22$ for GW170608 and GW190814, respectively \cite{Abbott:2020jks}.
The precession effect of GW190814 is well constrained to be small \cite{Abbott:2020khf}.
We use the public data and PSDs for the analysis \cite{190814data,190814psd}.

As a comparison, we also analyze the events with the IMRPhenomD waveform model, which is a frequency-domain aligned spin model including merger and ringdown parts as well \cite{Husa:2015iqa,Khan:2015jqa}. 

For the TaylorF2 waveform, we cutoff the waveform at $f_{\rm high} = 180$ Hz and $f_{\rm high} = 140$ Hz for GW170608 and GW190814, respectively, which roughly correspond to the boundary of the inspiral and the post inspiral parts for the corresponding event in the IMRPhenomD waveform model \cite{Abbott:2020jks}.
The lower cutoff frequency is $f_{\rm low} = 20$ Hz for GW170608.
For GW190814, we set  $f_{\rm low} = 20$ Hz for Handford and Virgo, and $f_{\rm low} = 30$ Hz for Livingston, following the analysis by LIGO and Virgo collaborations (LVC) \cite{Abbott:2020khf}.
For the IMRPhenomD waveform model, we choose $f_{\rm high} = 700$ Hz and $f_{\rm high} = 500$ Hz for GW170608 and GW190814, respectively, so that the post-inspiral part is also included.
We apply SS and BH priors for the TaylorF2 model and the BH prior for the IMRPhenomD model.

\begin{center}
\begin{figure*}[htbp]
\includegraphics[scale=0.56]{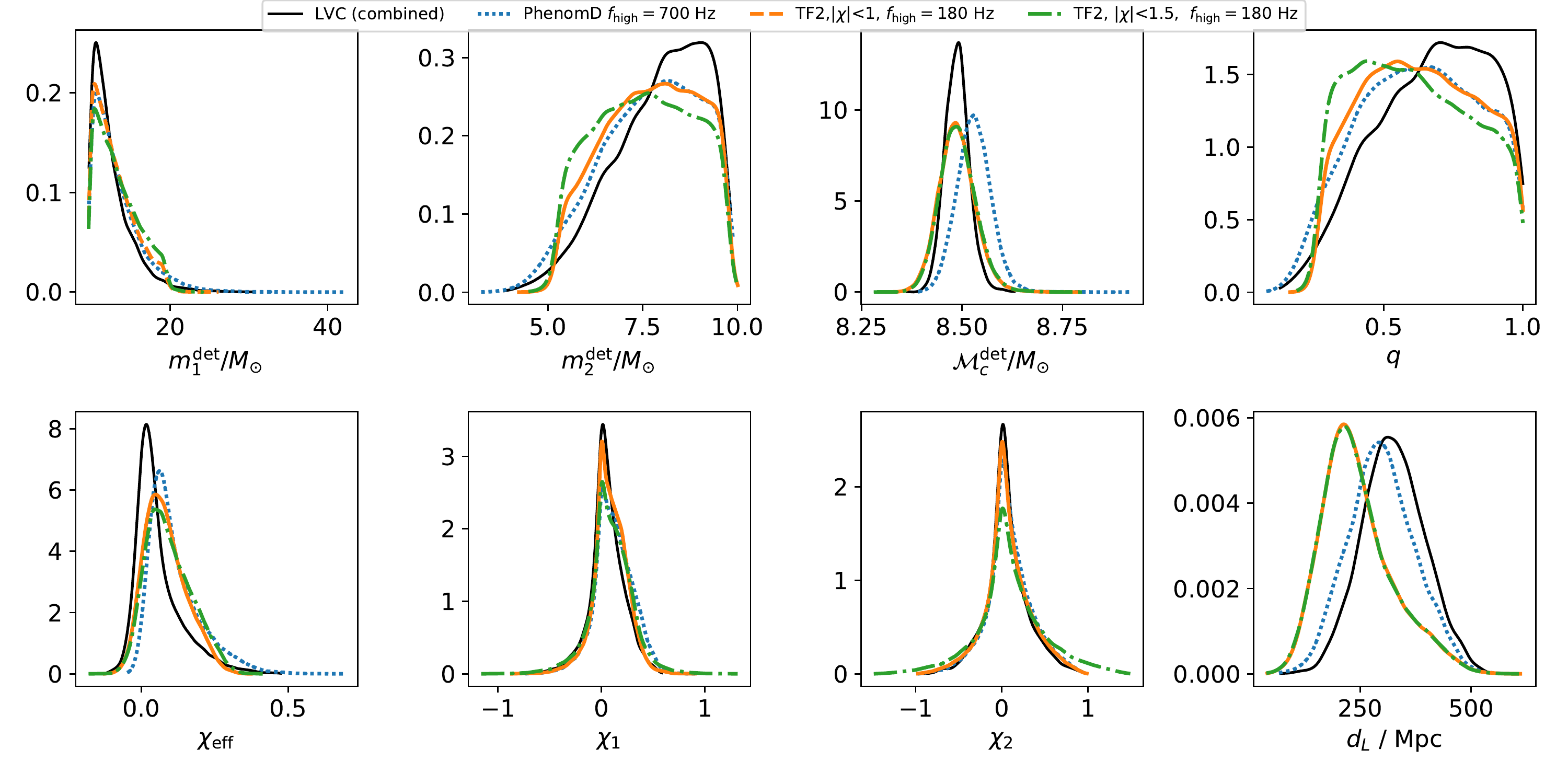}
\caption{The posterior distributions of mass parameters in the detector frame, spin parameters, and the luminosity distance $d_L$ for GW170608.
Blue, orange, and green curves correspond to IMRPhenomD ($f_{\rm high} = 700$Hz), TaylorF2 (BH), and TaylorF2 (SS).
Black curve is a posterior distribution from LVC public posterior samples, which is a combined one from the results of two waveform models (SEOBNRv3 and IMRPhenomPv2).}
\label{GW170608}
\end{figure*}
\end{center}

\begin{center}
\begin{figure*}[htbp]
\includegraphics[scale=0.56]{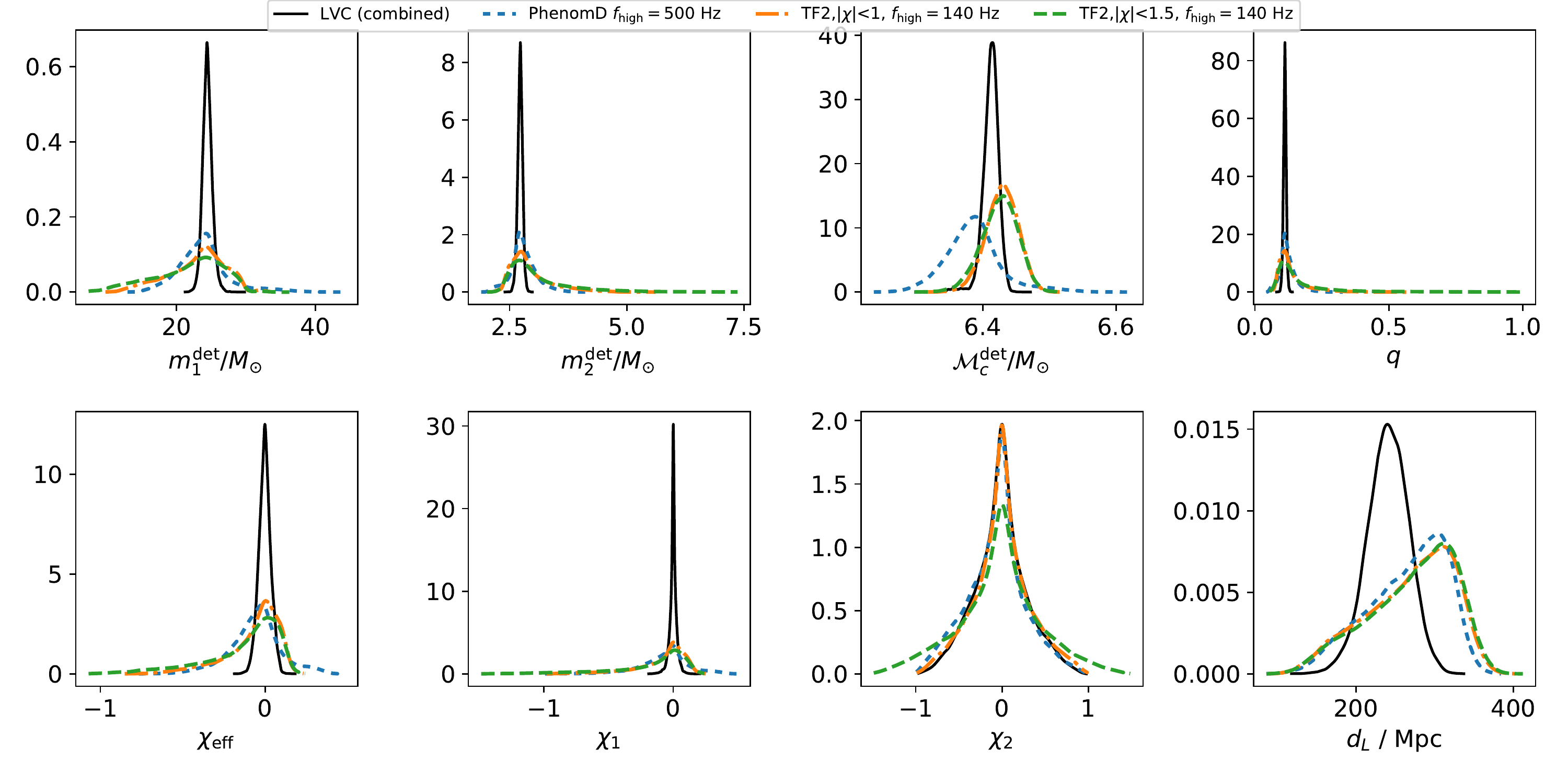}
\caption{Same as Fig.~\ref{GW170608}  but for GW190814.
Blue, orange, and green curves correspond to  IMRPhenomD ($f_{\rm high} = 500$Hz), TaylorF2 (BH), and TaylorF2 (SS).
Black curve is a posterior distribution from LVC public posterior samples, which is a combined one from the results of two waveform models (SEOBNRv4PHM and IMRPhenomPv3HM).}
\label{GW190814}
\end{figure*}
\end{center}

The posterior distributions of mass parameters in the detector frame, spin parameters and the luminosity distance are shown in Figs.~\ref{GW170608} and \ref{GW190814} for GW170608 and GW190814, respectively.
From Fig.~\ref{GW170608}, the posterior distributions of our results are consistent with the LVC results. 
Since we do not see much differences by comparing the results of IMRPhenomD and TaylorF2 (BH), we can assume that the posteriors of this event is well recovered by the inspiral only waveform.
From the results of TaylorF2 (BH) and TaylorF2 (SS), again we do not see much differences in both posteriors, only the tails of  the posteriors of $\chi_1$ and $\chi_2$ extend to $|\chi| >1$ for SS prior due to the wider spin prior range.
The posteriors of $\chi_1$ and $\chi_2$ almost represent the priors of those, although $\chi_{\rm eff}$ is well constrained to the positive and small value. 
Logarhithmic signal-to-noise Bayes factors for IMRPhenomD, TaylorF2 (BH) and TaylorF2 (SS) are $87.58, 78.53$ and $78.07$, respectively.

For GW190814 in Fig.~\ref{GW190814}, similar to GW170608, the posterior distributions of our results are consistent with the LVC results.  The larger statistical error arises from the lack of precessing effects and the higher multipole modes in the waveform model in our study \cite{Abbott:2020khf}.
Similar to GW170608 results, we do not see much differences in IMRPhenomD and TaylorF2 (BH), which implies the results can be recovered by the inspiral part.
By comparing TaylorF2 (BH) and TaylorF2 (SS), we do not see much difference in posterior distributions. 
For $\chi_{\rm eff}$ and $\chi_1$, the positive values are restricted to be small while the posterior tails extend to the negative value $\chi < -1$.
These tails might be caused by the lack of higher order modes or precession effect, since the tails do not exist in the LVC results.
For the secondary spin, the posterior represents the prior distribution, that is, we are not able to get any meaningful information.

Logarhithmic signal-to-noise Bayes factors for IMRPhenomD, TaylorF2 (BH) and TaylorF2 (SS) are $208.3, 207.1$ and $207.0$, respectively.

The 90\% symmetric credible regions of mass and spin parameters as well as the luminosity distance are summarized in Table \ref{credible_event} for TaylorF2 with the spin prior $|\chi_{1,2}| \le 1.5$. 
For GW170608, $\chi_{\rm eff}$ and $\chi_1$ are constrained to be smaller than $|\chi| < 0.5$ even for the spin prior $|\chi_{1,2}| \le 1.5$.
For GW190814, the positive bounds of these parameters are restricted to be smaller than $\chi = 0.12$, while the negative bounds are not tightly constrained. 
For both events, $\chi_2$ has a large statistical error, which does not give a meaningful constraint.
 
\begin{center}
\begin{table*}[htbp]
 \caption{The symmetric 90\% credible regions of $(m_1, m_2, {\mathcal M}_c, q, \chi_{\rm eff}, \chi_1, \chi_2, d_L)$ for TaylorF2 with the spin prior $|\chi| \le 1.5$. }
\begin{tabular}{l l l l l l l l l }
\hline \hline
Event &$m_1 / M_{\odot}$ & $m_2 / M_{\odot}$ & $\mathcal{M}_c^{\rm det}  / M_{\odot}$& $q$ &$\chi_{\rm eff}$ & $\chi_1$ & $\chi_2$ & $d_L$ /Mpc \\ \hline
GW170608 &$(10,18)$& $(5.5,9.5)$ & $(8.42,8.57)$ &$(0.30,0.95)$ &$(-0.01,0.24)$& $(-0.25,0.42)$ & $(-0.56,0.82)$ & $(131,402)$ \\ 
GW190814 &$(12,29)$& $(2.5,4.6)$ & $(6.38,6.47)$ & $(0.09,0.38)$ &$(-0.64,0.11)$& $(-0.82,0.12)$ & $(-0.89,0.79)$ & $ (166,345)$  \\
  \hline \hline
\end{tabular}
  \label{credible_event}
\end{table*}
\end{center}

\section{Summary and conclusion} 
We have analyzed inspiral gravitational waveforms from compact binaries, in which at least one component of the binary has $|\chi | >1$.
We apply two spin priors, $|\chi_{1,2}| \le 1$ and $|\chi_{1,2}| \le 1.5$, and investigate whether and how parameter estimation is affected by the spin prior $|\chi_{1,2}| \le 1$. 
We have found that when the primary is a superspinar, both mass and spin parameters are biased in parameter estimation due to the spin prior $|\chi_{1,2}| \le 1$.
All spin posteriors concentrate against the positive bound for this case.
We do not see a degeneracy between mass and spin parameters that can explain the bias in mass parameters due to the prior $|\chi_{1,2}| \le 1$.
The results from the prior $|\chi_{1,2}| \le 1.5$ shows a strong evidence compared to those from the prior $|\chi_{1,2}| \le 1$.

On the other hand, when the primary is a black hole, we do not see much bias in parameter estimation due to the limited spin prior range, even though the secondary is a superspinar.
We also obtain a weak support for $|\chi_{1,2}| \le 1.5$ prior for this case.

We conclude that the extension of the spin prior range is necessary for accurate parameter estimation if binaries with $\chi_{\rm eff} \approx 1$ are found, while it is difficult to identify superspinars if they are only the secondary objects.
Nevertheless, we may assume the objects as superspinars even when the spin prior is limited, if we see the spin posteriors are concentrating on the positive bound of the prior range as shown in Fig.~\ref{kde-a1_1.1-a2_1.1_q0.1}.

We also apply the analysis to black hole binary merger events GW170608 and GW190814, which have long and loud inspiral signals.
We do not see any preference of superspinars from the model selection for both events.

Since the TaylorF2 model becomes inappropriate in the late inspiral, improvements in the waveform model valid in $|\chi| >1$ are needed for more accurate parameter estimation for superspinars. 

In this study, we have assumed that we have detected superspinar binaries as black hole binaries in advance.
However, the template bank used for binary black holes search \cite{Abbott:2020niy} should be different from that for highly spinning binaries such as $\chi_{\rm eff} > 1$, it is important to reconstruct a template bank to detect superspinars.
As a future work, we will search for superspinar binaries with $\chi_{\rm eff} > 1$.

\begin{acknowledgements}
We would like to thank Soichiro Morisaki, Kyohei Kawaguchi, and Hideyuki Tagoshi for fruitful discussions.
We would also like to thank Takahiro Tanaka for reviewing the manuscript.
This research has made use of data, software, and web tools 
obtained from the Gravitational Wave Open Science
Center (https://www.gw-openscience.org), a service
of LIGO Laboratory, the LIGO Scientic Collaboration
and the Virgo Collaboration. LIGO is funded by the
U.~S.~National Science Foundation. Virgo is funded by
the French Centre National de la Recherche Scientique
(CNRS), the Italian Istituto Nazionale di Fisica Nucleare
(INFN), and the Dutch Nikhef, with contributions by
Polish and Hungarian institutes.
This research has also made use of Computing Infrastructure ORION in Osaka
City University. 
This work is supported by JSPS KAKENHI Grant Number JP17H06361 and JP21K03548.

\end{acknowledgements}
\appendix

\section{Effect of abrupt cutoff of the TaylorF2 waveform model}
\label{app2}
\begin{center}
\begin{figure*}[htbp]
\includegraphics[scale=0.56]{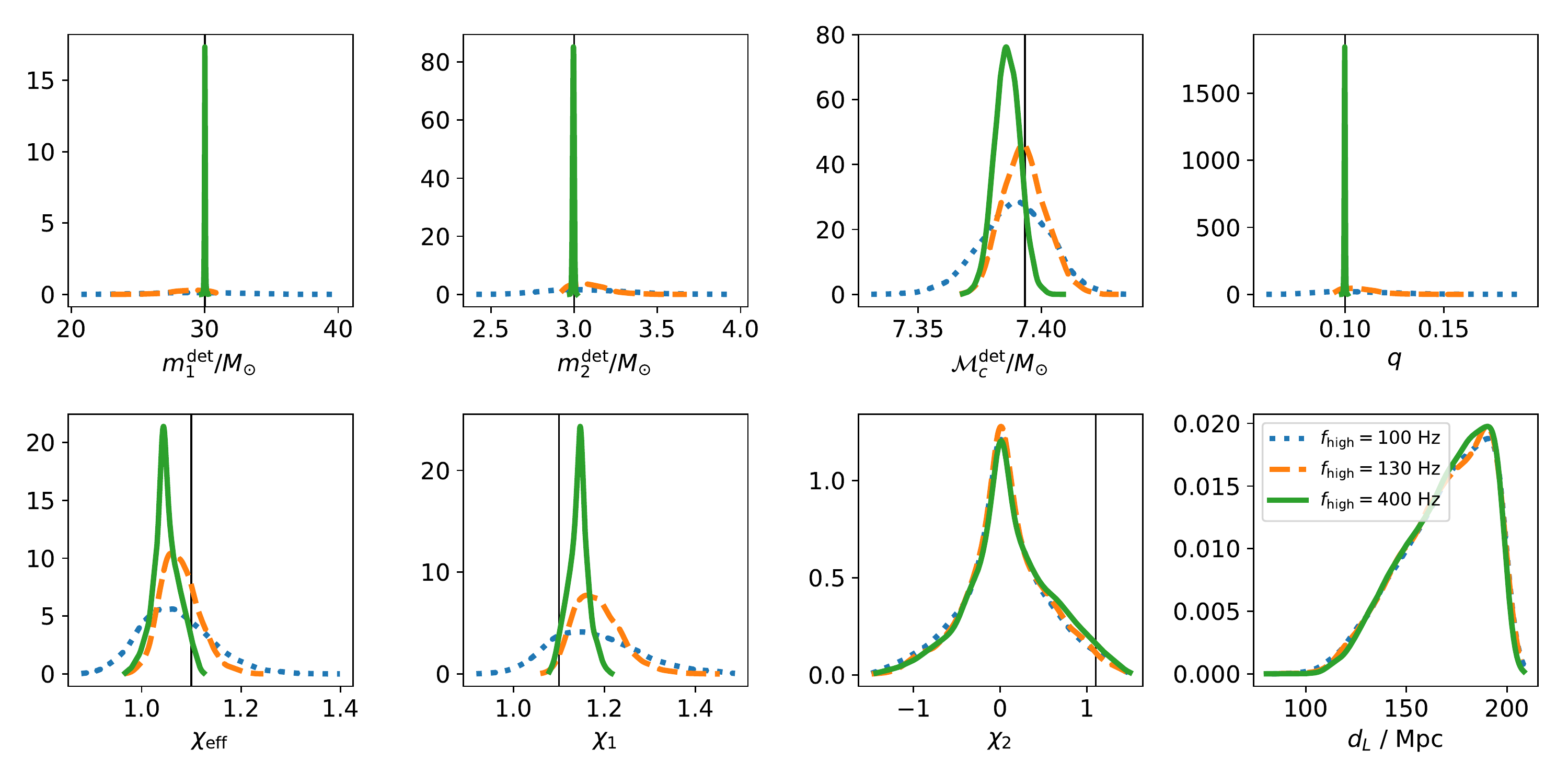}
\caption{The posterior distributions of mass parameters in the detector frame, spin parameters, and the luminosity distance $d_L$.
Injected values are $(m_1, m_2, \chi_1, \chi_2) = (30M_{\odot},3 M_{\odot} , 1.1, 1.1)$ with the spin prior $|\chi| \le 1.5$.
Blue,  orange, and green curves corresponds to $f_{\rm high}= 100$ Hz, $130$ Hz and $400$ Hz.
The vertical lines correspond to the injected values.}
\label{fhigh}
\end{figure*}
\end{center}

In this appendix, we discuss whether the abrupt cut-off of the TaylorF2 waveform model can affect the analysis given in the main text.
To investigate the effect, we compare the results of parameter estimation for three different $f_{\rm high}$.
Figure \ref{fhigh} shows the posterior distributions for the injected value of $(m_1, m_2, \chi_1, \chi_2) = (30M_{\odot},3 M_{\odot} , 1.1, 1.1)$ with the spin prior $|\chi| \le 1.5$ and 90\% symmetric credible regions are summarized in Table \ref{credible_fhigh}.
The injected waveform terminates at $\sim 133$ Hz.
For the mass parameters, $\chi_{\rm eff}$, and $\chi_1$, statistical errors are large for $f_{\rm high} < 133$ Hz, while $\chi_2$ and $d_L$ do not show any difference due to different $f_{\rm high} $.
We can see that $\mathcal{M}_c$ and $\chi_{\rm eff}$ for $f_{\rm high} = 400$ Hz have the smallest statistical errors compared to other $f_{\rm high}$ but peaks slightly deviate from the injected values.
These systematic biases might be caused from the abrupt cutoff of the waveform.

On the other hand, the posteriors of TaylorF2 show a consistency with those of IMRPhenomD for the real event analyses as shown in Sec.~IV.C.
Furthermore, the consistency of posteriors of the chirp mass between the TaylorF2 and a time domain inspiral-merger-ringdown waveform models is shown in Ref.~\cite{Sennett:2019bpc} for GW151226 and GW170608.

\begin{center}
\begin{table*}[htbp]
 \caption{The symmetric 90\% credible regions with respect to different $f_{\rm high}$ for the spin prior $|\chi| \le 1.5$. }
\begin{tabular}{l l l l l l l l l }
\hline \hline
  $f_{\rm high}$ &$m_1 / M_{\odot}$ & $m_2 / M_{\odot}$ & $\mathcal{M}_c^{\rm det}  / M_{\odot}$& $q$ &$\chi_{\rm eff}$ & $\chi_1$ & $\chi_2$ & $d_L$ /Mpc\\ \hline
100 Hz &$(24.3,34.3)$& $(2.71,3.51)$ & $(7.37,7.41)$ & $(0.08,0.14)$ &(0.96,1.20)& $(1.04,1.35)$ & $(-0.81,0.91)$& $(128,198)$\\ 
130 Hz &$(26.4,30.6)$& $(2.95,3.30)$   & $(7.38,7.41) $ &$(0.10,0.12)$ &(1.02,1.15)&  $(1.11,1.28)$ & $(-0.71,0.91)$ &$(130,197)$ \\
400 Hz &$(29.9,30.6) $& $(2.99,3.00)$& $ (7.38,7.39)$ &$ (0.10,0.10)$ &(1.01,1.09)& $(1.10,1.18)$ & $(-0.72,1.00) $ &$(131,197)$ \\ 
Injection &$30 $& $3$& $ 7.39$ &$0.1 $&$1.1$& $1.1$& $ 1.1$ &$200 $  \\  
  \hline \hline
\end{tabular}
   \label{credible_fhigh}
\end{table*}
\end{center}


\newpage
\begin{thebibliography}{99}


\bibitem{Carter:1971zc}
B.~Carter,
Phys. Rev. Lett. \textbf{26}, 331-333 (1971).

\bibitem{Gimon:2007ur}
E.~G.~Gimon and P.~Horava,
Phys. Lett. B \textbf{672}, 299-302 (2009).

\bibitem{Patil:2015fua}
M.~Patil, T.~Harada, K.~i.~Nakao, P.~S.~Joshi and M.~Kimura,
Phys. Rev. D \textbf{93}, no.10, 104015 (2016).

\bibitem{Cardoso:2008kj}
V.~Cardoso, P.~Pani, M.~Cadoni and M.~Cavaglia,
Class. Quant. Grav. \textbf{25}, 195010 (2008).

\bibitem{Pani:2010jz}
P.~Pani, E.~Barausse, E.~Berti and V.~Cardoso,
Phys. Rev. D \textbf{82}, 044009 (2010).

\bibitem{Nakao:2017rgv}
K.~i.~Nakao, P.~S.~Joshi, J.~Q.~Guo, P.~Kocherlakota, H.~Tagoshi, T.~Harada, M.~Patil and A.~Krolak,
Phys. Lett. B \textbf{780}, 410-413 (2018).


\bibitem{Roy:2019uuy}
R.~Roy, P.~Kocherlakota and P.~S.~Joshi,
[arXiv:1911.06169 [gr-qc]].


 \bibitem{Friedman:1978hf}
J.~L.~Friedman and B.~F.~Schutz,
Astrophys. J. \textbf{222}, 281 (1978).

\bibitem{Friedman:1978wla}
J.~L.~Friedman,
Commun. Math. Phys. \textbf{62}, no.3, 247-278 (1978).

\bibitem{Maggio:2018ivz}
E.~Maggio, V.~Cardoso, S.~R.~Dolan and P.~Pani,
Phys. Rev. D \textbf{99}, no.6, 064007 (2019).


\bibitem{TheLIGOScientific:2016wfe}
B.~P.~Abbott \textit{et al.} [LIGO Scientific and Virgo],
Phys. Rev. Lett. \textbf{116}, no.24, 241102 (2016).

\bibitem{170104}
B.~P.~Abbott \textit{et al.} [LIGO Scientific and VIRGO],
Phys. Rev. Lett. \textbf{118}, no.22, 221101 (2017)
[erratum: Phys. Rev. Lett. \textbf{121}, no.12, 129901 (2018)].

\bibitem{170814}
B.~P.~Abbott \textit{et al.} [LIGO Scientific and Virgo],
Phys. Rev. Lett. \textbf{119}, no.14, 141101 (2017).

\bibitem{170608}
B.~P.~Abbott \textit{et al.} [LIGO Scientific and Virgo],
Astrophys. J. \textbf{851}, no.2, L35 (2017).


\bibitem{gwtc-1}
B.~P.~Abbott {\it et al.} [LIGO Scientific and Virgo Collaborations],
  Phys.\ Rev.\ X {\bf 9}, 031040 (2019).
  
\bibitem{190412}
R.~Abbott \textit{et al.} [LIGO Scientific and Virgo],
Phys. Rev. D \textbf{102}, no.4, 043015 (2020).

  
  \bibitem{190521}
R.~Abbott \textit{et al.} [LIGO Scientific and Virgo],
Phys. Rev. Lett. \textbf{125}, no.10, 101102 (2020).
  
 \bibitem{Abbott:2020mjq}
R.~Abbott \textit{et al.} [LIGO Scientific and Virgo],
Astrophys. J. Lett. \textbf{900}, L13 (2020).


\bibitem{Abbott:2020niy}
R.~Abbott \textit{et al.} [LIGO Scientific and Virgo],
[arXiv:2010.14527 [gr-qc]].


\bibitem{Biscoveanu:2020are}
S.~Biscoveanu, M.~Isi, S.~Vitale and V.~Varma,
[arXiv:2007.09156 [astro-ph.HE]].

\bibitem{Mandel:2020lhv}
I.~Mandel and T.~Fragos,
Astrophys. J. Lett. \textbf{895}, no.2, L28 (2020).

\bibitem{Zevin:2020gxf}
M.~Zevin, C.~P.~L.~Berry, S.~Coughlin, K.~Chatziioannou and S.~Vitale,
Astrophys. J. Lett. \textbf{899}, no.1, L17 (2020).

\bibitem{Dhurandhar:1992mw}
S.~V.~Dhurandhar and B.~S.~Sathyaprakash,
Phys. Rev. D \textbf{49}, 1707-1722 (1994).

\bibitem{Buonanno:2009zt}
A.~Buonanno, B.~Iyer, E.~Ochsner, Y.~Pan and B.~S.~Sathyaprakash,
Phys. Rev. D \textbf{80}, 084043 (2009).

 \bibitem{Blanchet:2013haa}
L.~Blanchet,
Living Rev. Rel. \textbf{17}, 2 (2014).

 \bibitem{Khan:2015jqa}
S.~Khan, S.~Husa, M.~Hannam, F.~Ohme, M.~P\"urrer, X.~Jim\'enez Forteza and A.~Boh\'e,
Phys. Rev. D \textbf{93}, no.4, 044007 (2016).
 
 \bibitem{VanDenBroeck:2006ar}
C.~Van Den Broeck and A.~S.~Sengupta,
Class. Quant. Grav. \textbf{24}, 1089-1114 (2007).

\bibitem{Wade:2013hoa}
M.~Wade, J.~D.~E.~Creighton, E.~Ochsner and A.~B.~Nielsen,
Phys. Rev. D \textbf{88}, no.8, 083002 (2013).

\bibitem{Piovano:2020ooe}
G.~A.~Piovano, A.~Maselli and P.~Pani,
[arXiv:2003.08448 [gr-qc]].

\bibitem{Piovano:2020zin}
G.~A.~Piovano, A.~Maselli and P.~Pani,
[arXiv:2004.02654 [gr-qc]].

 
\bibitem{Abbott:2020khf}
R.~Abbott \textit{et al.} [LIGO Scientific and Virgo],
Astrophys. J. Lett. \textbf{896}, no.2, L44 (2020).

\bibitem{Abbott:2020jks}
R.~Abbott \textit{et al.} [LIGO Scientific and Virgo],
[arXiv:2010.14529 [gr-qc]].



\bibitem{Bambi:2011mj}
C.~Bambi,
Mod. Phys. Lett. A \textbf{26}, 2453-2468 (2011).


\bibitem{Bardeen:1972fi}
J.~M.~Bardeen, W.~H.~Press and S.~A.~Teukolsky,
Astrophys. J. \textbf{178}, 347 (1972).

\bibitem{Bambi:2008jg}
C.~Bambi and K.~Freese,
Phys. Rev. D \textbf{79}, 043002 (2009).
[

\bibitem{Zhou:2019kwb}
B.~Zhou, A.~Tripathi, A.~B.~Abdikamalov, D.~Ayzenberg, C.~Bambi, S.~Nampalliwar and M.~Zhou,
Eur. Phys. J. C \textbf{80}, no.5, 400 (2020).

\bibitem{Veitch:2014wba}
J.~Veitch, V.~Raymond, B.~Farr, W.~Farr, P.~Graff, S.~Vitale, B.~Aylott, K.~Blackburn, N.~Christensen, M.~Coughlin, W.~Del Pozzo, F.~Feroz, J.~Gair, C.~J.~Haster, V.~Kalogera, T.~Littenberg, I.~Mandel, R.~O'Shaughnessy, M.~Pitkin, C.~Rodriguez, C.~R\"over, T.~Sidery, R.~Smith, M.~Van Der Sluys, A.~Vecchio, W.~Vousden and L.~Wade,
Phys. Rev. D \textbf{91}, no.4, 042003 (2015).

\bibitem{lal}
LIGO Scientic Collaboration, {\it LIGO Algorithm
Library - LALSuite}, Free Software (GPL), 2018;
\texttt{https://doi.org/10.7935/GT1W-FZ16}.

\bibitem{skilling}
J. Skilling, Bayesian Analysis {\bf 1}, 833 (2006).

\bibitem{Veitch:2009hd}
J.~Veitch and A.~Vecchio,
Phys. Rev. D \textbf{81}, 062003 (2010).




 \bibitem{Lange:2018pyp}
J.~Lange, R.~O'Shaughnessy and M.~Rizzo,
[arXiv:1805.10457 [gr-qc]].

\bibitem{TheLIGOScientific:2014jea}
J.~Aasi \textit{et al.} [LIGO Scientific],
Class. Quant. Grav. \textbf{32}, 074001 (2015).

\bibitem{ligopsd}
\url{https://dcc.ligo.org/public/0002/T0900288/003/AdvLIGO%20noise%20curves.pdf}

\bibitem{TheVirgo:2014hva}
F.~Acernese \textit{et al.} [VIRGO],
Class. Quant. Grav. \textbf{32}, no.2, 024001 (2015).

\bibitem{Mandel:2014tca}
I.~Mandel, C.~P.~L.~Berry, F.~Ohme, S.~Fairhurst and W.~M.~Farr,
Class. Quant. Grav. \textbf{31}, 155005 (2014).

\bibitem{Ajith:2009bn}
P.~Ajith, M.~Hannam, S.~Husa, Y.~Chen, B.~Bruegmann, N.~Dorband, D.~Muller, F.~Ohme, D.~Pollney and C.~Reisswig, \textit{et al.}
Phys. Rev. Lett. \textbf{106}, 241101 (2011).

 \bibitem{Purrer:2015nkh}
M.~P\"urrer, M.~Hannam and F.~Ohme,
Phys. Rev. D \textbf{93}, no.8, 084042 (2016).

 
\bibitem{Chatziioannou:2018wqx}
K.~Chatziioannou, G.~Lovelace, M.~Boyle, M.~Giesler, D.~A.~Hemberger, R.~Katebi, L.~E.~Kidder, H.~P.~Pfeiffer, M.~A.~Scheel and B.~Szil\'agyi,
Phys. Rev. D \textbf{98}, no.4, 044028 (2018).

\bibitem{170817}
B.~P.~Abbott \textit{et al.} [LIGO Scientific and Virgo],
Phys. Rev. X \textbf{9}, no.1, 011001 (2019).

\bibitem{jeffreys}
H.~Jeffreys,
{\it The Theory of Probability}, (Oxford England, Oxford 1961), 3rd edn..

\bibitem{190814data}
\url{ https://www.gw-openscience.org/eventapi}

\bibitem{190814psd}
\url{ https://dcc.ligo.org/LIGO-P2000223/public}

\bibitem{Husa:2015iqa}
S.~Husa, S.~Khan, M.~Hannam, M.~P\"urrer, F.~Ohme, X.~Jim\'enez Forteza and A.~Boh\'e,
Phys. Rev. D \textbf{93}, no.4, 044006 (2016).



\bibitem{Sennett:2019bpc}
N.~Sennett, R.~Brito, A.~Buonanno, V.~Gorbenko and L.~Senatore,
Phys. Rev. D \textbf{102}, no.4, 044056 (2020).

 




 

 \end{thebibliography}
\end{document}